\documentclass[10pt,twoside,twocolumn,english,superscriptaddress,notitlepage,aps,prl]{revtex4-2}

\usepackage[LGR,T1]{fontenc}
\usepackage[utf8]{inputenc}
\usepackage[letterpaper]{geometry}
\usepackage{xcolor}
\usepackage{pdfcolmk}
\usepackage{babel}
\usepackage{verbatim}
\usepackage{textcomp}
\usepackage{amsbsy}
\usepackage{amstext}
\usepackage{graphicx}
\usepackage{esint}
\PassOptionsToPackage{normalem}{ulem}
\usepackage{ulem}
\usepackage[unicode=true,pdfusetitle,
 bookmarks=false,
 breaklinks=true,pdfborder={0 0 0},pdfborderstyle={},backref=false,colorlinks=true]
 {hyperref}
\hypersetup{
 colorlinks=true,citecolor=blue,linkcolor=black,urlcolor=black,pagebackref=true}

\usepackage{float} 
\usepackage{amssymb} 
\usepackage{gensymb} 

\makeatletter

\usepackage{amsfonts,anyfontsize,xcolor,graphicx}
\usepackage[calcwidth,explicit]{titlesec}
\usepackage[normalem]{ulem}

\titleformat{\section}{\bfseries\large\sffamily\scshape\filcenter}{\thesection.}{0.2em}{#1}  
\titlespacing{\section}{0pt}{0.2ex}{0.2ex}
\titleformat{\subsection}{\bfseries\normalsize\sffamily}{}{0em}{#1} 
\titlespacing{\subsection}{0pt}{0.5ex}{0ex}
\titleformat{\paragraph}[runin]{\normalfont\normalsize\bfseries}{}{0pt}{\theparagraph.}
\titlespacing*{\paragraph}{0em}{0ex}{0.3em}[]

\makeatletter\renewcommand\frontmatter@abstractwidth{\dimexpr\textwidth-2cm\relax}\makeatother

\setcitestyle{square,super}

\renewcommand\thesection{\Alph{section}}
\renewcommand{\theparagraph}{\Alph{section}\arabic{paragraph}}
\makeatletter\@addtoreset{paragraph}{section}\makeatother
\makeatletter\def\p@paragraph{}\makeatother

\addto\captionsenglish{\renewcommand{\figurename}{Figure}}

\AtBeginDocument{
\renewcommand{\ref}[1]{\autoref{#1}}
\renewcommand{\figureautorefname}{Figure}
\renewcommand{\equationautorefname}{Equation}
\renewcommand{\tableautorefname}{Table}
\renewcommand{\sectionautorefname}{\S}
\renewcommand{\subsectionautorefname}{\S}
\renewcommand{\paragraphautorefname}{\S}
}

\makeatother

\begin{document}

\title{Unveiling the emergent traits of chiral spin textures in magnetic multilayers}

\author{Xiaoye Chen}
\affiliation{Institute of Materials Research \& Engineering, Agency for Science,
Technology \& Research (A{*}STAR), 138634 Singapore}
\affiliation{Data Storage Institute, Agency for Science, Technology \& Research
(A{*}STAR), 138634 Singapore}

\author{Ming Lin}
\thanks{These authors contributed equally to this work.}
\affiliation{Institute of Materials Research \& Engineering, Agency for Science,
Technology \& Research (A{*}STAR), 138634 Singapore}

\author{Jian Feng Kong}
\thanks{These authors contributed equally to this work.}
\affiliation{Institute of High Performance Computing, Agency for Science,
Technology \& Research (A{*}STAR), 138632 Singapore}

\author{Hui Ru Tan}
\affiliation{Institute of Materials Research \& Engineering, Agency for Science,
Technology \& Research (A{*}STAR), 138634 Singapore}

\author{Anthony K.C. Tan}
\affiliation{Data Storage Institute, Agency for Science, Technology \& Research
(A{*}STAR), 138634 Singapore}

\author{Soong-Geun Je}
\affiliation{Center for X-ray Optics, Lawrence Berkeley National Laboratory, Berkeley, California 94720, USA }

\author{Hang Khume Tan}
\affiliation{Institute of Materials Research \& Engineering, Agency for Science,
Technology \& Research (A{*}STAR), 138634 Singapore}
\affiliation{Data Storage Institute, Agency for Science, Technology \& Research
(A{*}STAR), 138634 Singapore}

\author{Khoong Hong Khoo}
\affiliation{Institute of High Performance Computing, Agency for Science,
Technology \& Research (A{*}STAR), 138632 Singapore}

\author{Mi-Young Im}
\affiliation{Center for X-ray Optics, Lawrence Berkeley National Laboratory, Berkeley, California 94720, USA }

\author{Anjan Soumyanarayanan}
\email{anjan@imre.a-star.edu.sg}
\affiliation{Institute of Materials Research \& Engineering, Agency for Science,
Technology \& Research (A{*}STAR), 138634 Singapore}
\affiliation{Data Storage Institute, Agency for Science, Technology \& Research
(A{*}STAR), 138634 Singapore}
\affiliation{Physics Department, National University of Singapore (NUS), 117551
Singapore }

\begin{abstract}
Magnetic skyrmions are topologically wound nanoscale textures of spins whose ambient stability and electrical manipulation in multilayer films have led to an explosion of research activities. While past efforts focused predominantly on isolated skyrmions, recently ensembles of chiral spin textures, consisting of skyrmions and magnetic stripes, were shown to possess rich interactions with potential for device applications. However, several fundamental aspects of chiral spin texture phenomenology remain to be elucidated, including their domain wall structure, thermodynamic stability, and morphological transitions. Here we unveil the evolution of these textural characteristics on a tunable multilayer platform -- wherein chiral interactions governing spin texture energetics can be widely varied -- using a combination of full-field electron and soft X-ray microscopies with numerical simulations. With increasing chiral interactions, we demonstrate the emergence of Néel helicity, followed by a marked reduction in domain compressibility, and finally a transformation in the skyrmion formation mechanism. Together with an analytical model, these experiments establish a comprehensive microscopic framework for investigating and tailoring chiral spin texture character in multilayer films.
\end{abstract}

\maketitle

\section*{Introduction\label{sec:intro}}


Seminal advances in tailoring interfacial interactions in magnetic thin films have led to the room temperature (RT) stabilization of nanoscale spin textures -- most notably magnetic skyrmions \citep{moreau-luchaire_additive_2016, woo_observation_2016, boulle_room-temperature_2016, soumyanarayanan_tunable_2017}. In light of past efforts on magnetic bubbles and domain walls (DWs) \citep{bobeck_magnetic_1975, parkin_magnetic_2008}, the excitement around magnetic skyrmions stems largely from their non-trivial topology, small size, and their coupling to electrical stimuli. Firstly, skyrmions possess finite topological charge, which emerges from the material-specific handedness and manifests as the chirality of spins winding around their centre\cite{nagaosa_topological_2013}. The unique spin structure of skyrmions facilitates their electrical detection\cite{neubauer_topological_2009, nagaosa_topological_2013}, while also enabling them to remain stable at sizes down to 2~nm \citep{romming_writing_2013, hagemeister_stability_2015}. Importantly, they can be electrically generated, \citep{jiang_blowing_2015, buttner_field-free_2017, woo_deterministic_2018, je_targeted_2021} and driven at relatively high efficiency using electrical currents \citep{woo_observation_2016, reichhardt_collective_2015}.

While such sparse, isolated skyrmions in chiral multilayers are attractive for spintronic device applications, equally ripe for exploitation are denser \textbf{\textit{ensembles of chiral spin textures}}, consisting of magnetic stripes and skyrmions. Several recent proposals look to harness these so-called ``skyrmion fabrics''   \citep{prychynenko_magnetic_2018, pinna_reservoir_2020}  for reservoir computing. However, a comprehensive microscopic picture of chiral spin texture phenomenology, and their response to external stimuli, remains to be established. While some reports confirm their Néel helicity in chiral multilayers\citep{boulle_room-temperature_2016, mcvitie_transmission_2018}, others have reported a considerable Bloch component with a layer- and material-dependent magnitude \citep{legrand_hybrid_2018, dovzhenko_magnetostatic_2018}. Meanwhile, experimental investigations of the field evolution of their size have focused narrowly on ``bubble skyrmions'' \citep{boulle_room-temperature_2016}. Even theoretical efforts, while fully exploring isolated skyrmions, are yet to examine the rich interactions between skyrmions and stripes \citep{buttner_theory_2018, bernand-mantel_skyrmion-bubble_2018}. Finally, while morphological transitions from stripes to skyrmions have been demonstrated \citep{lemesh_current-induced_2018, je_targeted_2021}, the thermodynamic mechanism of skyrmion formation is not understood. Elucidating the microscopic origin of these attributes necessitates a multi-modal investigation of chiral spin textures with varying magnetic interactions on a single material platform.


Much of the character of chiral spin textures can be tuned by a single material parameter, $\kappa=\pi D/4\sqrt{AK_\text{eff}}$, where $D$ is the interfacial Dzyaloshinskii-Moriya interaction (iDMI), $A$ is the exchange stiffness and $K_\text{eff}$ is the effective uniaxial anisotropy \citep{bogdanov_thermodynamically_1994, rohart_skyrmion_2013, leonov_properties_2016}. Within a simple analytical model without assumption of sample symmetry, $\kappa$ is associated with the DW energy --- $\kappa > 1$ implies that DW energy density is negative \citep{bogdanov_thermodynamically_1994, legrand_modeling_2018}.  Here, we investigate the emergent  characteristics of chiral spin textures over a wide range of $\kappa$ on a tunable Co/Pt-based multilayer platform. Exploiting the complementary sensitivity of Lorentz transmission electron microscopy (LTEM) and magnetic transmission soft X-ray microscopy (MTXM), to textural characteristics, we elucidate the evolution of DW helicity, domain compressibility, and skyrmion formation mechanism with increasing $\kappa$.  In conjunction with micromagnetic simulations and an analytical model, we establish a microscopic framework for spin texture character in multilayer films.

Our work is performed at RT using Co/Pt-based multilayer stacks with out-of-plane (OP) anisotropy, which are established hosts of magnetic textures \citep{belliard_stripe_1997}. While symmetric stacks have negligible total iDMI, asymmetric stacks, such as (Ir or Ta)/Co/Pt can have sizable iDMI ($D>1$~mJ/m$^2$) -- relevant to chiral magnetic textures \citep{moreau-luchaire_additive_2016, woo_observation_2016, boulle_room-temperature_2016}.  The inclusion of Fe -- as in Ir/Fe/Co/Pt stacks -- enhances the iDMI, while $D$ and $K_\text{eff}$ can be smoothly modulated by Fe and Co thicknesses \citep{soumyanarayanan_tunable_2017}. Here we study four samples each comprising 1~nm thick FM layers -- identified by their \textbf{Fe(x)/Co(y)} composition (\ref{tab:table1}) -- wherein the active stack is repeated 14 times to optimize full-field magnetic contrast. Interfacial interactions are progressively introduced and quantified using established techniques \citep{moreau-luchaire_additive_2016, woo_observation_2016, soumyanarayanan_tunable_2017}, with the estimated iDMI ($D_\text{est}$) varying over 0 -- 2~mJ/m$^2$ and $K_\text{eff}$ over 0.08 -- 0.70~MJ/m$^3$ (Experimental Section, SM1). Notably, the $D_\text{est}$ determined for the 14-stack multilayers studied here are in line with measured values on corresponding single stacks from Brillouin light scattering (BLS) experiments  \citep{bottcher_dzyaloshinskii-moriya_2020}. Consequently, $\kappa$ varies over 0 -- 1.5, and provides the requisite range for mapping magnetic texture evolution. 

\begin{table}[b]

\begin{ruledtabular}
\begin{tabular}{llccc}

\textbf{Acronym}&
\textbf{Stack Composition (\r{A})}&
\textbf{$\boldsymbol{K}_\text{eff}$}&
\textbf{$\boldsymbol{D}_\text{est}$}&
\textbf{$\boldsymbol\kappa$}\\
\colrule
\textbf{$^\text{S}$Co(10)} & [Pt(10)/Co(10)/Pt(10)]$_{14}$ & 0.68 & 0 & 0\\
{\footnotesize\textbf{Fe(0)/Co(10)}} & [Ir(10)/Co(10)/Pt(10)]$_{14}$ & 0.47 & 1.3 & 0.3\\
\textbf{Fe(2)/Co(8)} & {\footnotesize[Ir(10)/Fe(2)/Co(8)/Pt(10)]$_{14}$} & 0.22 & 1.8 & 0.9\\
\textbf{Fe(3)/Co(7)} & {\footnotesize [Ir(10)/Fe(3)/Co(7)/Pt(10)]$_{14}$} & 0.08 & 2.0 & 1.5\\
\end{tabular}
\end{ruledtabular}

\caption{\label{tab:table1}%
\textbf{Sample Compositions.}
List of multilayer samples used in this work, with layer thickness in angstroms in parentheses (see Experimental Section, SM1 for full-stack details). Corresponding magnetic properties are listed: effective anisotropy \textbf{$K_\text{eff}$} (MJ/m$^3$), estimated iDMI \textbf{$D_\text{est}$} (mJ/m$^2$) and the stability parameter, \textbf{$\kappa$}. The samples are henceforth referred to by their \textbf{acronym}. }
\end{table}

\section*{DW Helicity\label{sec:helicity}}

\begin{figure}
\includegraphics[width=\columnwidth]{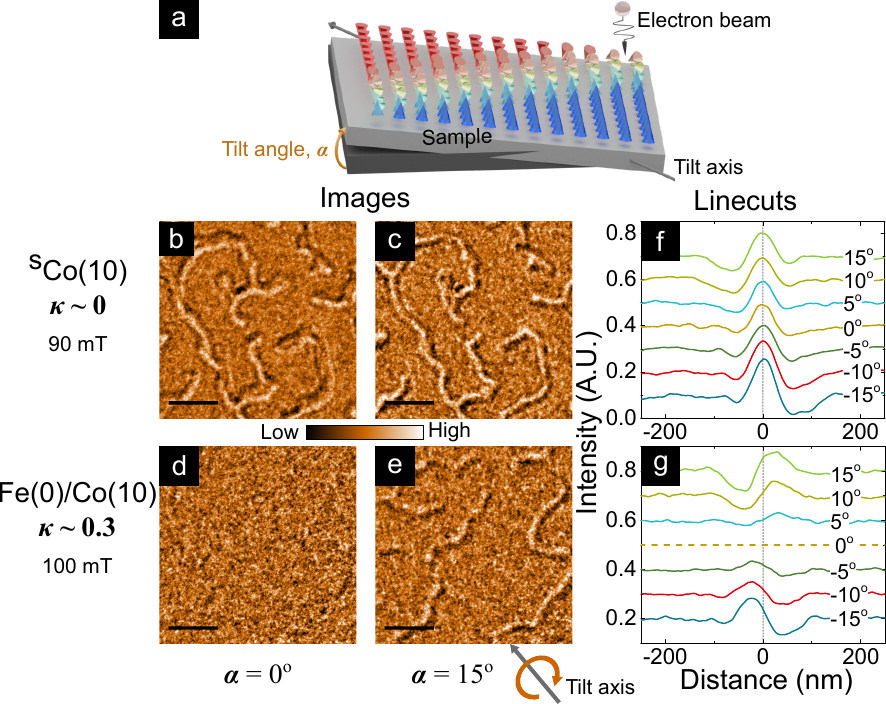}
\caption{\label{fig:helicity} \textbf{DW Helicity from Tilt-Dependent LTEM Imaging.} \textbf{(a)} Schematic diagram of LTEM imaging geometry with sample (illustrated here with a Néel DW) tilted at angle $\alpha$ with respect to the plane normal to the electron beam. The tilt axis is shown below (e). \textbf{(b-e)} Representative LTEM images (scale bar: 0.5~$\mu$m) acquired on samples $^\text{S}$Co(10) (b-c) and Fe(0)/Co(10) (d-e) at $\mu_0 H$ = 90 and 100~mT respectively for $\alpha=0\degree$ (b,d) and $\alpha=15\degree$ (c,e), with $-2$~mm defocus. \textbf{(f-g)} Average cross-sectional linecuts across domains detected in LTEM images of $^\text{S}$Co(10) (f) and Fe(0)/Co(10) (g), with $\alpha$ varied over $\pm15\degree$. Each curve represents the average of $\sim 1,000$ linecuts extracted from domains imaged over a 5~$\mu$m field-of-view using an automated recipe. Dashed vertical lines mark the domain center.}
\end{figure}

The introduction of iDMI should lead to a measurable change in DW helicity of skyrmion textures \citep{nagaosa_topological_2013, hellman_interface-induced_2017}. LTEM imaging -- wherein magnetic contrast results from the magnetization curl parallel to the electron beam -- is particularly sensitive to such changes. For normal beam incidence (zero tilt), a pair of homochiral Bloch DWs should express symmetric contrast about their center, while Néel DWs should exhibit no contrast as their curl is perpendicular to the beam \citep{benitez_magnetic_2015}. Meanwhile, the positions of Néel DWs can be deduced by tilting the sample (\ref{fig:helicity}a), whereupon antisymmetric domain contrast can be observed \citep{mcvitie_transmission_2018}. To visualize this evolution, we perform tilt-dependent LTEM imaging with samples deposited on SiO$_x$ membranes (see Experimental Section). For ease of analysis, we use OP magnetic fields ($\mu_0 H$) large enough to ensure adjacency of pairs of DWs (i.e. thin domains). Artifacts due to granularity and membrane waviness are mitigated using a recipe that extracts $\sim$1000 linecuts across domains imaged over a 5~$\mu$m field-of-view (see SM2). 

\begin{figure*}[t]
\includegraphics[width=2\columnwidth]{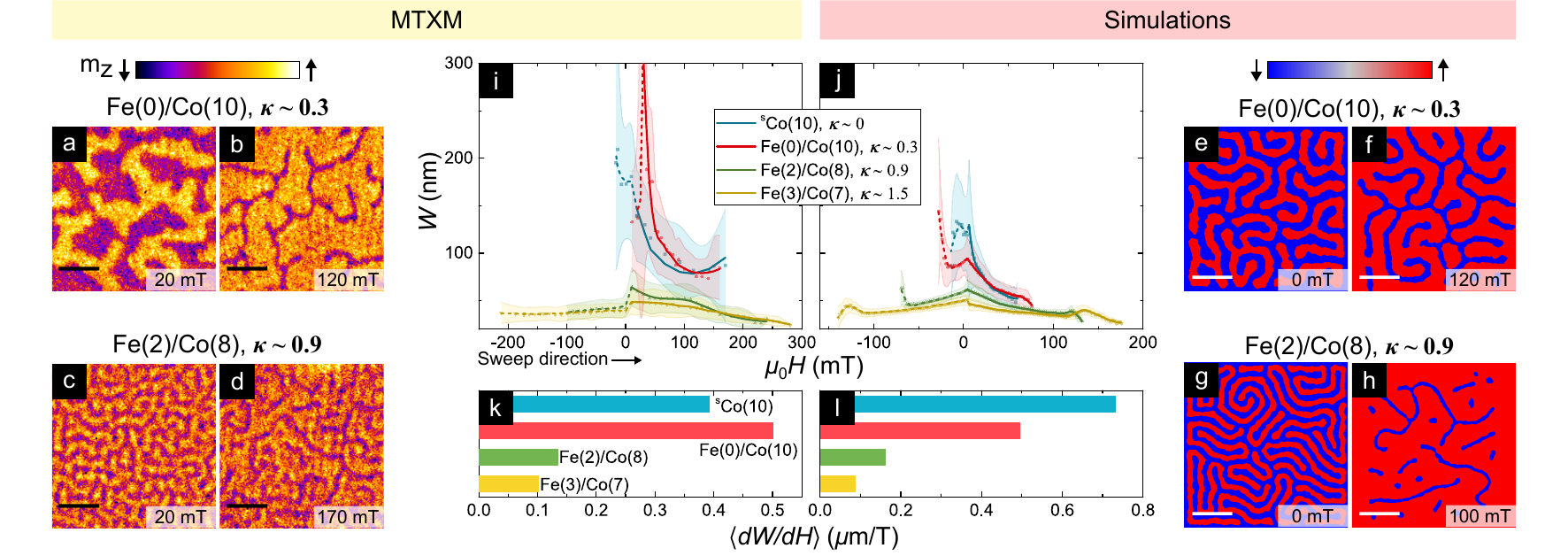}
\caption{\label{fig:compressibility} \textbf{Domain Width Field Evolution: MTXM Imaging and Simulations.} \textbf{(a-d)} MTXM images of samples Fe(0)/Co(10) ($\kappa \sim 0.3$: a-b) and Fe(2)/Co(8) ($\kappa \sim 0.9$, c-d) -- showing domain evolution from near-zero (a,c) to near-saturation (b,d) fields. (scalebar: 0.5~$\mu$m) \textbf{(e-h)} Simulated magnetization images for magnetic parameters consistent with Fe(0)/Co(10) (e-f) and Fe(2)/Co(8) (g-h) -- showing the corresponding evolution from zero (e,g) to higher fields (f,h). \textbf{(i-j)} Average minority polarisation domain width, $W(H)$, with varying magnetic field from MTXM experiments (i) and simulations (j) on all four samples studied in this work.  Each $W(H)$ data point is a full field-of-view mean (shaded band is the standard deviation) determined using an automated recipe (see SM2). Solid and dashed lines, guides to the eye, represent magnetization in the negative and positive vertical direction respectively.  \textbf{(k-l)} Average magnitude of compressibility, $\langle dW/dH \rangle$, for the four samples from experiments (k) and simulations (l), corresponding to the average gradients in (i) and (j) respectively  for $\mu_0 H>0$. }
\end{figure*}

\ref{fig:helicity} shows stark differences in tilt-dependent LTEM results for $^\text{S}$Co(10) and Fe(0)/Co(10). First, $^\text{S}$Co(10) ($D_\text{est}\simeq0$, $\kappa\simeq0$) shows strong, symmetric contrast about the domain center at zero tilt (\ref{fig:helicity}b), with a small antisymmetric component at finite tilt angle (\ref{fig:helicity}c,f). This is consistent with Bloch DWs expected for symmetric stacks \citep{chess_determination_2017, garlow_quantification_2019, fallon_quantitative_2019}. Micromagnetic simulations performed with $^\text{S}$Co(10) parameters (see Experimental Section) suggest that the Bloch DWs are achiral, i.e. lack fixed handedness (see SM4). In comparison, Fe(0)/Co(10) ($D_\text{est} \simeq 1.3$~mJ/m$^2$, $\kappa \simeq 0.3$) shows no contrast at zero tilt (\ref{fig:helicity}d). Contrast at finite tilt is consistently antisymmetric -- and whose amplitude increases with tilt angle (\ref{fig:helicity}e,g) -- consistent with Néel DWs. The lack of measurable symmetric contrast in \ref{fig:helicity}g suggests that any Bloch component -- e.g. due to layer-dependent chirality \citep{legrand_hybrid_2018, dovzhenko_magnetostatic_2018} -- is negligibly small \citep{fallon_quantitative_2019}. Micromagnetic simulations for Fe(0)/Co(10) also reflect the limited influence of such layer dependent variations, which are further suppressed if moderate interlayer exchange coupling is included (see SM4). Finally, similar experiments on $\kappa \gtrsim 1$ samples produce results fully consistent with Fe(0)/Co(10) (see SM3). These results indicate that chiral interactions in the $\kappa \sim 0.3$ sample are sufficiently large to transform achiral Bloch textures ($\kappa \simeq 0$) to homochiral Néel textures\citep{legrand_modeling_2018}. 

\section*{Domain Compressibility}

Having established DW helicity evolution, we turn to domain characteristics -- which evolve with OP field in addition to magnetic interactions \citep{woo_observation_2016, moreau-luchaire_additive_2016, soumyanarayanan_tunable_2017}. Both stripes and skyrmions can be collectively characterized by a single length scale, $W$, defined as domain width of stripes and diameter of skyrmions of the minority polarisation. Notably, the field-induced variation of $W$, or $\boldsymbol{dW}/\boldsymbol{dH}$ -- termed as \textbf{\textit{domain compressibility}} \citep{vernier_measurement_2014} -- should also evolve with $\kappa$ \citep{rohart_skyrmion_2013, soumyanarayanan_tunable_2017, ho_geometrically_2019}. MTXM imaging -- wherein magnetic circular dichroic contrast is proportional to local OP magnetization \citep{fischer_magnetic_2013} -- is well-suited to measure $W$. Therefore, we performed MTXM imaging with varying OP field using samples deposited on Si$_3$N$_4$ membranes,  complemented by micromagnetic simulations (see Experimental Section). $W$ was determined as an averaged quantity over the full field-of-view of about 5~$\mu$m using an automated recipe to mitigate granularity effects (see SM2). The identity of the minority polarisation flips at the coercive field, resulting in a sharp kink in $W$.

\ref{fig:compressibility} shows MTXM (a-d,i,k) and simulation results (e-h,j,l) of $W(H)$ across samples with varying $\kappa$. On one hand, for $\kappa \ll 1$ -- illustrated for Fe(0)/Co(10) ($\kappa \sim 0.3$, \ref{fig:compressibility}a-b,i) -- $W$ shrinks rapidly with field ($\langle dW/dH \rangle \sim 0.5$~$\mu$m/T, \ref{fig:compressibility}k). Such \textbf{\textit{highly compressible behavior}} is well reproduced by simulations (\ref{fig:compressibility}e-f). One difference from experiments is the relative order of $\langle dW/dH \rangle$ for $^\text{S}$Co(10) and Fe(0)/Co(10). The likely source of this discrepancy is the domain nucleation field, which affects domain compressibility. In low $\kappa$ samples such as $^\text{S}$Co(10) and Fe(0)/Co(10), domain nucleation may be dominated by extrinsic factors such as grains \citep{quinteros_impact_2020} that are not accounted for in our simulations. Nevertheless, the $W(H)$ trend of chiral Néel textures (Fe(0)/Co(10)) is remarkably similar to achiral Bloch textures ($^\text{S}$Co(10)) \citep{montoya_tailoring_2017}, \ref{fig:compressibility}i-j). This suggests that domain compressibility is largely independent of DW helicity. On the other hand, for $\kappa \gtrsim 1$ -- shown for Fe(2)/Co(8) ($\kappa \sim 0.9$, \ref{fig:compressibility}c-d,i) --  the $W(H)$ variation is much reduced  ($\langle dW/dH \rangle \sim 0.1$~$\mu$m/T, \ref{fig:compressibility}k) \citep{romming_field-dependent_2015}. Similarly rigid or incompressible behavior is seen for Fe(3)/Co(7) ($\kappa \sim 1.5$), albeit at reduced $W$, and in the analysis of LTEM images (see SM3). Finally, simulated trends for $\kappa \gtrsim 1$ are also in line with these results (\ref{fig:compressibility}g-h,j,l), suggesting that the contrast may be understood within a micromagnetic energy framework. 



To elucidate the compressibility evolution, we use an analytical model of 1D periodic domains within an infinite magnetic slab of thickness $t$, domain period $\lambda$, and DW width $\Delta$ (Figure 3a). This model is chosen because it considers the interactions between neighbouring stripes, which we believe is essential to the compressibility phenomenon. The total energy density is given by \citep{buttner_theory_2018}:
\begin{widetext}
\begin{equation}\label{eqn:etotal}
    \varepsilon_{\text{tot}} = \frac{2}{\lambda}\left[ \frac{2A}{\Delta} + 2K_u\Delta + \pi D\sin\psi \right] + \varepsilon_{d,s} + \varepsilon_{d,v}- M_s \left( 1-\frac{2W}{\lambda} \right) B_z,
\end{equation}
where the magnetostatic energy densities due to surface ($\varepsilon_{d,s}$)  and volume charges ($\varepsilon_{d,v}$) are: 
\begin{equation}\label{eqn:eds}
    \varepsilon_{d,s} = \frac{\mu_0 M_s^2}{2} \left( 1-\frac{2W}{\lambda}\right)^2 + \frac{2\pi\mu_0 M_s^2 \Delta^2}{\lambda t} \sum_{n=1}^\infty \frac{\sin^2\frac{\pi n W}{\lambda}}{\sinh^2\frac{\pi^2 n\Delta}{\lambda}}\frac{1-\exp\left(-\frac{2\pi n t}{\lambda}\right)}{n}, \text{and}
\end{equation}
\begin{equation}\label{eqn:evs}
    \varepsilon_{d,v} = \frac{2\pi\mu_0 M_s^2 \Delta^2 \sin^2\psi}{\lambda t} \sum_{n=1}^\infty \frac{\sin^2\frac{\pi n W}{\lambda}}{\cosh^2\frac{\pi^2 n\Delta}{\lambda}} \frac{\exp\left(-\frac{2\pi n t}{\lambda}\right) + \frac{2\pi n t}{\lambda} - 1}{n}.
\end{equation}
\end{widetext}

\begin{figure}[t]
\includegraphics[width=\columnwidth]{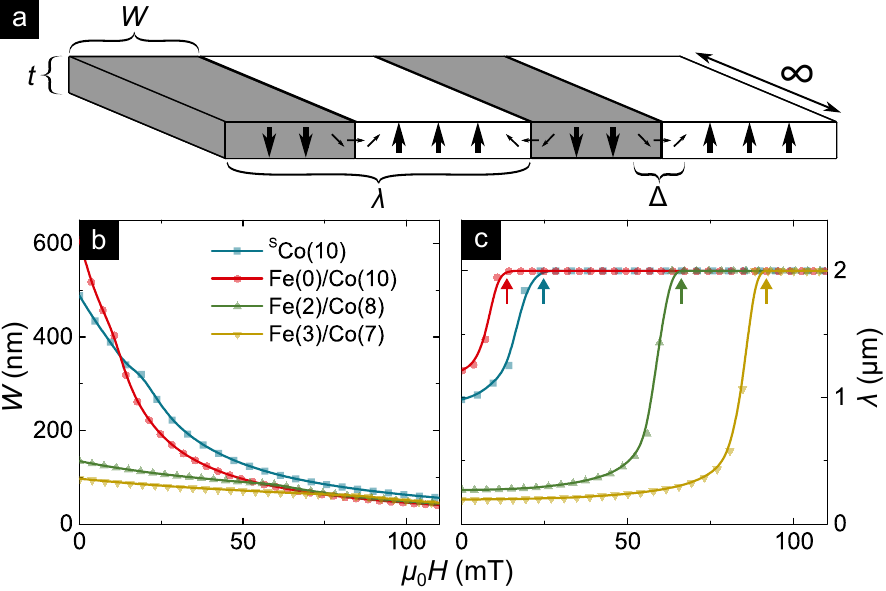}
\caption{\label{fig:model} \textbf{1D Model for Domain Compressibility Evolution.} \textbf{(a)} Schematic of the simplified analytical model of 1D periodic domains used to interpret the observed domain compressibility. Minority domains of DW width $\Delta$, total width $W$, and period $\lambda$ are considered within an infinitely long magnetic slab of thickness $t$ and breadth 2~$\mu$m. \textbf{(b-c)} Field evolution of domain width, $W(H)$ (b, c.f. \ref{fig:compressibility}i-j) and period, $\lambda(H)$ (c), obtained from the 1D model (details in text) for magnetic parameters consistent with the samples of interest. Arrows in (c) mark the saturation field, $H_S$, predicted by the model. }
\end{figure}

The field evolutions of $W$ and $\lambda$ (\ref{fig:model}b-c) are obtained by numerically minimizing $\varepsilon_\text{tot}$ with respect to $\lambda$, $W$, and $\Delta$. Notably, the 1D model reproduces $\kappa$ dependence of compressibility found in experiments (\ref{fig:compressibility}i) and simulations (\ref{fig:compressibility}j): domains are highly compressible for $\kappa \ll 1$, and relatively incompressible for $\kappa \gtrsim 1$.  Furthermore, it offers a physical explanation for the compressibility evolution when viewed in conjunction with $\lambda(H)$. The latter is indicative of the saturation field, $H_S$ (arrows in \ref{fig:model}c), and the domain density. For $\kappa \gtrsim 1$, wherein $H_S$ is higher (see SM1), domain nucleation occurs just below $H_S$ with smaller size (\ref{fig:model}b) and very close proximity (\ref{fig:model}c). The latter ensures  \textbf{\textit{mutual confinement}} of domains, limiting the expansion of $W$ with reducing $H$. Therefore, as $H$ is increased from zero, $\kappa \gtrsim 1$ domains have limited latitude for compression, and $W(H)$ is nearly constant -- expectedly near the lower cut-off ($W\sim \Delta$). The converse argument holds for $\kappa \ll 1$ domains, which explains their highly compressible $W(H)$ behavior.


The marked variation of domain compressibility with $\kappa$, its direct experimental accessibility, and consistency with grain-free simulations and the 1D model, establish compressibility as an important classifier of skyrmions (and stripes). Compressibility incorporates energetic considerations underlying a theoretically proposed “minimum skyrmion size” metric for isolated skyrmions\citep{buttner_theory_2018} (details in SM6), while also being relatively robust to material complexities such as grains and defects compared to the size of isolated skyrmions \citep{buttner_theory_2018, gross_skyrmion_2018}. Therefore, it can serve as a useful means to experimentally differentiate highly compressible “bubble” skyrmions from relatively incompressible “compact” skyrmions. Meanwhile, the remarkable difference in compressibility between samples with  $\kappa \ll 1$ and $\kappa \gtrsim 1$ demonstrates the importance of considering interactions between skyrmion textures within theoretical models. Further, it hints at the possibility of using effective fields, generated by material, geometric, or external means \citep{chen_room_2015,legrand_room-temperature_2019, ho_geometrically_2019}, to tune the size and morphology of stabilized spin textures.

\section*{Skyrmion Formation Mechanism}
\begin{figure*}
\includegraphics[width=2\columnwidth]{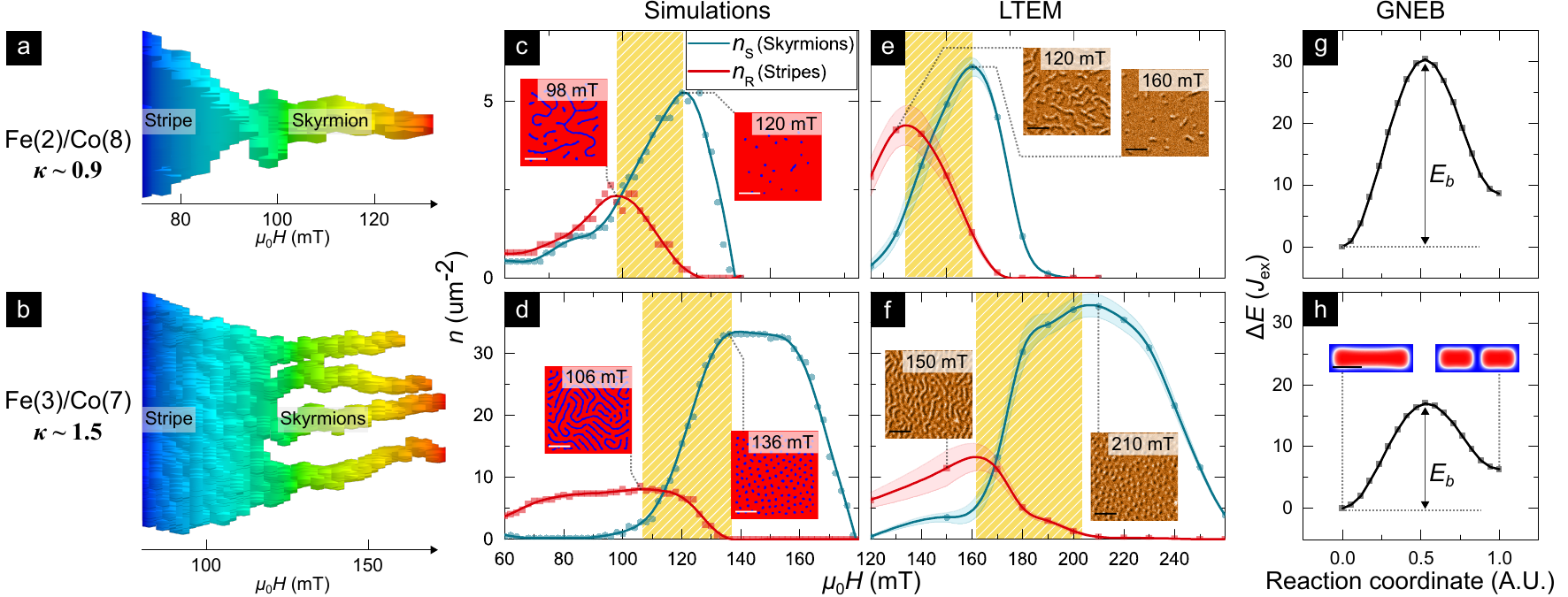}
\caption{\label{fig:formation} \textbf{Evolution of Skyrmion Formation Mechanism.} \textbf{(a-b)} Simulated field evolution of a prototypical stripe after its separation from the labyrinthine state for (a) Fe(2)/Co(8) ($\kappa \sim 0.9$) and (b) Fe(3)/Co(7) ($\kappa \sim 1.5$) parameters. The stripe is identified and isolated in real-space for each field slice and stacked horizontally to form a 3D structure, which is in turn reoriented with the length of the stripe along the vertical and  field axis along the horizontal (details in SM4). \textbf{(c-f)} Field evolution of the density of skyrmions ($n_S$, teal) and stripes ($n_R$, red), as extracted from simulations (c-d) and LTEM imaging (e-f) for samples Fe(2)/Co(8) (c,e) and Fe(3)/Co(7) respectively (procedural details in SM2). Highlighted regions denote field ranges corresponding to marked stripe to skyrmion transitions (c.f. a-b). Inset: Simulation and LTEM images immediately before and after stripe-to-skyrmion transitions (scalebar: 0.5~$\mu$m). \textbf{(g-h)} Energy profile (in units of Heisenberg exchange  $J_\text{ex}$) governing the fission of a stripe for Fe(2)/Co(8) ($\kappa \sim 0.9$) and Fe(3)/Co(7) ($\kappa \sim 1.5$) respectively calculated using GNEB method (details in SM5). $E_b$ denotes the energy barrier for the fission process, inset of (h) depicts the stripe before and after fission (scalebar: 0.1~$\mu$m).}
\end{figure*}

While skyrmions are known to emerge from stripes with increasing field, the transition may involve one or more mechanisms or paths.  Notably, $\kappa$, which determines DW stability, is also expected to affect this stripe-to-skyrmion transition. First, we visually examine the $\kappa$-variation of this transition by tracking the simulated field evolution of a prototypical magnetic stripe  (details in SM4). We see for $\kappa \sim 0.9$ (\ref{fig:formation}a) that the stripe shrinks smoothly with field, and eventually turns into a single skyrmion. Meanwhile, for $\kappa \sim 1.5$ (\ref{fig:formation}b), the stripe abruptly fissions into 4 distinct skyrmions at a characteristic field \citep{tan_skyrmion_2020}. These two mechanisms should result in contrasting textural field evolutions that should be detectable in our experiments. Therefore, we statistically examine the field evolution of stripes and skyrmions - distinguished in images by their circularity (see SM2). Here, we choose LTEM imaging, as it enables a clearer distinction between skyrmions and stripes (see SM2). 

\ref{fig:formation}c-f present the field evolution of densities of skyrmions ($n_S$) and stripes ($n_R$) from LTEM and simulations for samples Fe(2)/Co(8) and Fe(3)/Co(7). For each case, highlighted regions at intermediate fields -- spanning from  $n_R$ peak to $n_S$ peak -- indicate stripe-to-skyrmion transitions, and exhibit contrasting trends. For Fe(2)/Co(8) ($\kappa \sim 0.9$, \ref{fig:formation}c,e) the decrease in $n_R$ ($\sim$2--3~$\mu$m$^{-2}$) corresponds to a one-to-one increase in $n_S$ ($\sim$2--3~$\mu$m$^{-2}$). This is consistent with the shrinking of one stripe to one skyrmion, thereby resulting in isolated skyrmions (\ref{fig:formation}c,e: inset). In contrast, for Fe(3)/Co(7) ($\kappa \sim 1.5$, \ref{fig:formation}d,f) the decrease in $n_R$ ($\sim$7-8~$\mu$m$^{-2}$) coincides with a four-fold increase in $n_S$ ($\sim$30~$\mu$m$^{-2}$). This is in line with the fission of one stripe into $\sim4$ skyrmions on average and generates a dense skyrmion lattice (\ref{fig:formation}d,f: inset) \citep{tan_skyrmion_2020}. Thus, we have empirically observed the increased favorability of fission with increasing $\kappa$ (0.9 to 1.5).

The above observation may be understood from kinetic considerations. The fission of a stripe involves a change in topology and hence should be protected by an energy barrier. To examine the evolution of the barrier height ($E_b$) with $\kappa$, we perform geodesic nudged elastic band (GNEB) calculations for Fe(2)/Co(8) and Fe(3)/Co(7) (\ref{fig:formation}g,h, details in SM5). We found that $E_b$ for Fe(2)/Co(8) is 40\% greater than that for Fe(3)/Co(7). Assuming that entropic effects are comparable across the two compositions  \citep{desplat_thermal_2018}, it follows that fission will be greatly suppressed in Fe(2)/Co(8) relative to Fe(3)/Co(7). The suppression of fission in Fe(2)/Co(8) will then require stripes to instead smoothly shrink into skyrmions.  


\section*{Outlook}
\begin{figure}[b]
\includegraphics[width=\columnwidth]{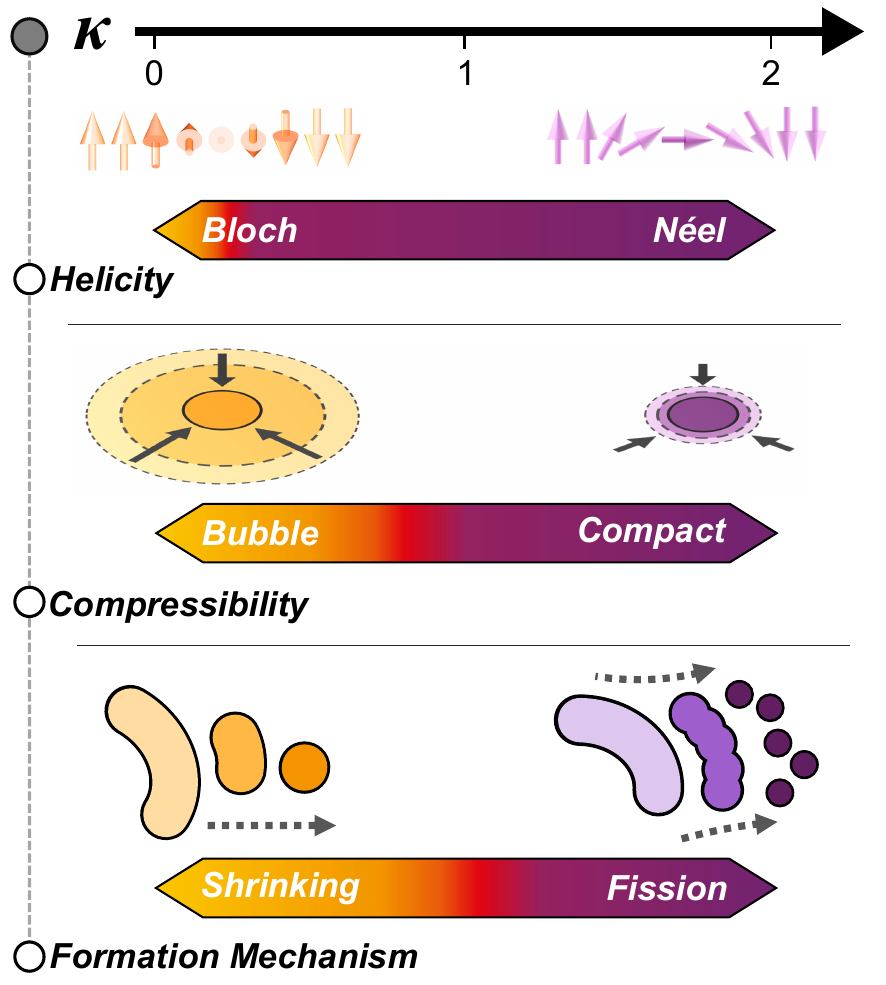}
\caption{\label{fig:summary} \textbf{$\kappa$-Driven Evolution of Skyrmion Character.} Overview of the evolution of multilayer skyrmion characteristics with increasing $\kappa$ as seen across the samples studied in this work. This includes the change of DW helicity from Bloch to Néel (top), domain compressibility from bubble to compact (middle), and skyrmion formation mechanism from shrinking to fission of stripes (bottom).}
\end{figure}


In summary, we have elucidated transitions in three critical characteristics of chiral spin textures. As shown in \ref{fig:summary}, these characteristics systematically evolve with $\kappa$ -- the material parameter determining chiral DW stability. First, as $\kappa$ increases measurably from zero, the DW helicity transitions from achiral Bloch to chiral Néel-type. Next, as $\kappa$ approaches unity, the domain compressibility is drastically reduced, transforming ``bubble'' skyrmions into ``compact'' skyrmions. Finally, for $\kappa > 1$, the skyrmion formation mechanism evolves from  shrinking to fission of stripes, resulting in the proliferation of compact Néel skyrmions for $\kappa > 1$. 


Our findings -- established on a single tunable material platform -- provide several valuable insights towards understanding the observed phenomenology of chiral spin textures, notably skyrmions. Firstly, we have shown that a small but finite $\kappa$ ($\sim 0.3$) enables the formation of Néel DWs with fixed chirality, with no evidence of a Bloch component even for a 14-repeat stack. Next, we have established the compressibility of domains as a robust experimental metric to differentiate bubble and compact skyrmions. Finally, we have shown that the preference for one of two distinct skyrmion formation mechanisms -- shrinking and fission -- may explain the observation of isolated skyrmions \citep{moreau-luchaire_additive_2016, woo_deterministic_2018} in some materials and dense skyrmion lattices \citep{romming_writing_2013,soumyanarayanan_tunable_2017} in others.

These insights provide a timely roadmap to inform stack design for skyrmionic applications -- particularly in device architectures that rely on ensembles of chiral spin textures rather than on sparse, isolated skyrmions. For example, selecting a stack with $0 < \kappa \ll 1$, hosting highly compressible domains, will enable dynamic tuning of the spin texture morphology with temporal variation of applied fields. Conversely, if the application requires control of the topology of textures, a $\kappa \gg 1$ stack, enabling fission-driven skyrmion formation, would be a better fit. Spanning the physics of stripes and skyrmions, our work provides a springboard for their use as ``skyrmion fabrics'' for applications in unconventional computing \citep{prychynenko_magnetic_2018}.

\section*{Methods\label{sec:Methods}}

\small{
\subsection*{Sample Fabrication}

Multilayer films, comprising Ta(40)/Pt(50)/\textbf{[HM(10)/Fe(x)/Co(y)/HM(10)]14}/Pt(20) (HM: heavy metal, number in parentheses indicates thickness in angstroms),  were deposited by DC magnetron sputtering at room temperature using a Chiron™ UHV system manufactured by Bestec GmbH (base pressure: $10^{-8}$ Torr). Four samples were studied in this work whose active stack compositions (bolded above) are listed in \ref{tab:table1}. To enable direct comparison between different techniques used in this work, the films were simultaneously deposited for magnetometry on thermally oxidized 100~nm Si wafer substrates, for LTEM on 20~nm-thick SiO$_2$ membrane window grids from SPI Supplies, and for MTXM on 50 -- 200~nm thick Si$_3$N$_4$ membranes from Silson. Magnetometry measurements were performed using an EZ11 vibrating sample magnetometer (VSM) made by MicroSense™. The magnetic parameters: $M_S$, $K_\text{eff}$, $A_\text{est}$ and $D_\text{est}$ were obtained using protocols consistent with literature \citep{woo_observation_2016, moreau-luchaire_additive_2016, soumyanarayanan_tunable_2017, ho_geometrically_2019, jin_control_2017}, and are detailed in SM1.

\subsection*{Lorentz TEM Experiments}

Lorentz transmission electron microscopy (LTEM) experiments were performed using an FEI Titan S/TEM operated in Lorentz Fresnel mode at 300~kV. A dedicated Lorentz lens for focusing the electron beam was used at a defocus of $-2$~mm. Meanwhile, the objective lens located at the sample position was switched off for imaging acquisition under field-free conditions, or excited to different strengths to apply out-of-plane magnetic fields ($-300$~mT to +2~T) for in-situ studies of magnetic texture evolution. 

\subsection*{MTXM Experiments}

Full-field MTXM imaging experiments were performed using circularly polarised soft x-rays at the Advanced Light Source (XM-1 BL 6.1.2), using the Co L3 edge ($\sim$778~eV) with out-of-plane (OP) sample geometry \citep{fischer_magnetic_2013}. OP magnetic fields were applied using an electromagnet, and a pair of horse-shoe poles were used to guide the generated flux. 

\subsection*{Micromagnetic Simulations}

Micromagnetic simulations were performed using MuMax3 to interpret the field evolution of the 14 repeat multilayer stacks \citep{vansteenkiste_design_2014}. The simulation field-of-view used was 2~$\mu$m $\times$ 2~$\mu$m, and the cell size was kept to 4~nm $\times$ 4~nm $\times$ 3~nm, which is below the exchange length for all samples. The effective medium approximation was used with one layer per stack repetition to account for memory constraints \citep{woo_observation_2016}. Hysteresis loops were simulated using protocols described in \citep{vansteenkiste_design_2014}.

\subsection*{Image Analysis}

Custom-written Python scripts were used for the quantitative analysis of magnetic microscopy images. These scripts comprise routines for image filtering and binarization followed by domain characterization and statistics using standard methods in the scikit-image library \citep{walt_scikit-image:_2014}. The analysis procedures are detailed in SM2.

}

\noindent \begin{center}
{\small{}\rule[0.5ex]{0.4\columnwidth}{0.5pt}}{\small\par}
\par\end{center}

\noindent {\small{}We acknowledge the support of the National Supercomputing Centre (NSCC) and A*STAR Computational Resource Centre (A*CRC) for computational resources. This work was supported
by the SpOT-LITE program (Grant Nos. A1818g0042, A18A6b0057), funded by Singapore's RIE2020 initiatives, and by the Pharos skyrmion program
(Grant No. 1527400026) funded by A{*}STAR, Singapore. Works at the ALS were supported by U.S. Department of Energy (DE-AC02-05CH11231). M.-Y. Im acknowledges support by Lawrence Berkeley National Laboratory through the Laboratory Directed Research and Development (LDRD) Program. The authors thank Pin Ho and Ramu Maddu for helpful discussions.}\textsf{\textbf{\small{}}}%

\phantomsection\addcontentsline{toc}{section}{\refname}\bibliography{Textures}


\normalsize
\clearpage
\onecolumngrid
\begin{center}
\textbf{\Large Supplementary Materials}
\end{center}
\setcounter{section}{0}
\setcounter{figure}{0}
\setcounter{table}{0}
\setcounter{equation}{0}

\setcounter{secnumdepth}{4}
\renewcommand\thesection{S\arabic{section}}
\renewcommand{\theparagraph}{S\arabic{section}\alph{paragraph}}
\makeatletter\@addtoreset{paragraph}{section}\makeatother
\makeatletter\def\p@paragraph{}\makeatother
\renewcommand{\thefigure}{S\arabic{figure}}
\renewcommand{\theequation}{S\arabic{equation}}
\renewcommand{\thetable}{S\arabic{table}}

\addto\captionsenglish{\renewcommand{\figurename}{Figure}}

\renewcommand{\ref}[1]{\autoref{#1}}
\renewcommand{\figureautorefname}{Figure}
\renewcommand{\equationautorefname}{Equation}
\renewcommand{\tableautorefname}{Table}
\renewcommand{\sectionautorefname}{\S}
\renewcommand{\subsectionautorefname}{\S}
\renewcommand{\paragraphautorefname}{\S}

\setcounter{tocdepth}{1}
\makeatletter\def\l@section{\@dottedtocline{1}{0.6em}{1.5em}}\makeatother
\makeatletter\def\l@paragraph{\@dottedtocline{4}{1.5em}{1.8em}}\makeatother
\makeatletter\def\l@figure{\@dottedtocline{1}{0.6em}{1.8em}}\makeatother

\linespread{1.25}
\def\arraystretch{1.5}
\setlength{\parskip}{1.5ex plus0.2ex minus0.2ex} 
\setlength{\abovecaptionskip}{4pt}\setlength{\belowcaptionskip}{-4pt}
\setlength{\abovedisplayskip}{0ex}\setlength{\belowdisplayskip}{0ex}
\setlength{\abovedisplayshortskip}{0ex}\setlength{\belowdisplayshortskip}{0ex}

\titleformat{\section}{\large\bfseries\scshape\filcenter}{\thesection.}{1em}{#1}[{\titlerule[0.5pt]}]
\titlespacing*{\section}{0pt}{1ex}{1ex}
\titleformat{\subsection}{\bfseries\sffamily}{\thessubsection.}{0em}{#1}
\titlespacing{\subsection}{0pt}{0.5ex}{0.5ex}
\titleformat{\paragraph}[runin]{\sffamily\bfseries}{}{-1.2em}{#1.}
\titlespacing*{\paragraph}{1.25em}{2ex}{0.4em}[] 

\section{Magnetic Properties of Multilayer Samples}

\begin{figure}[b]
\includegraphics[width=0.4\columnwidth]{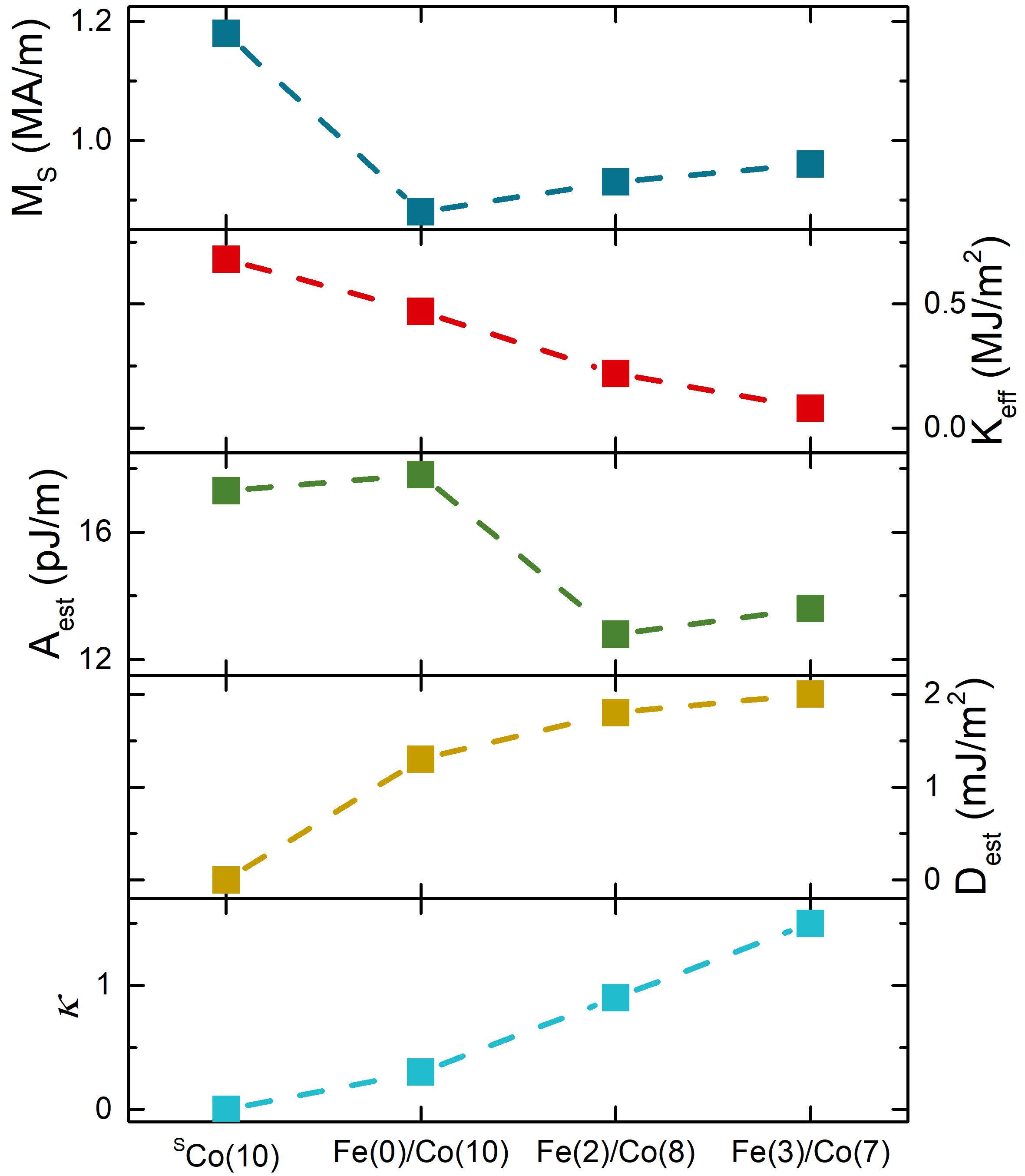}
\caption{\label{fig:magnetic_props} 
\textbf{Magnetic properties of multilayer samples.} 
The trendlines of $M_\text{s}$, $K_\text{eff}$, $A_\text{est}$, $D_\text{est}$ and $\kappa$ across samples. The samples are arranged in order of increasing $\kappa$.}
\end{figure}

\begin{figure}[b]
\includegraphics[width=0.896\columnwidth]{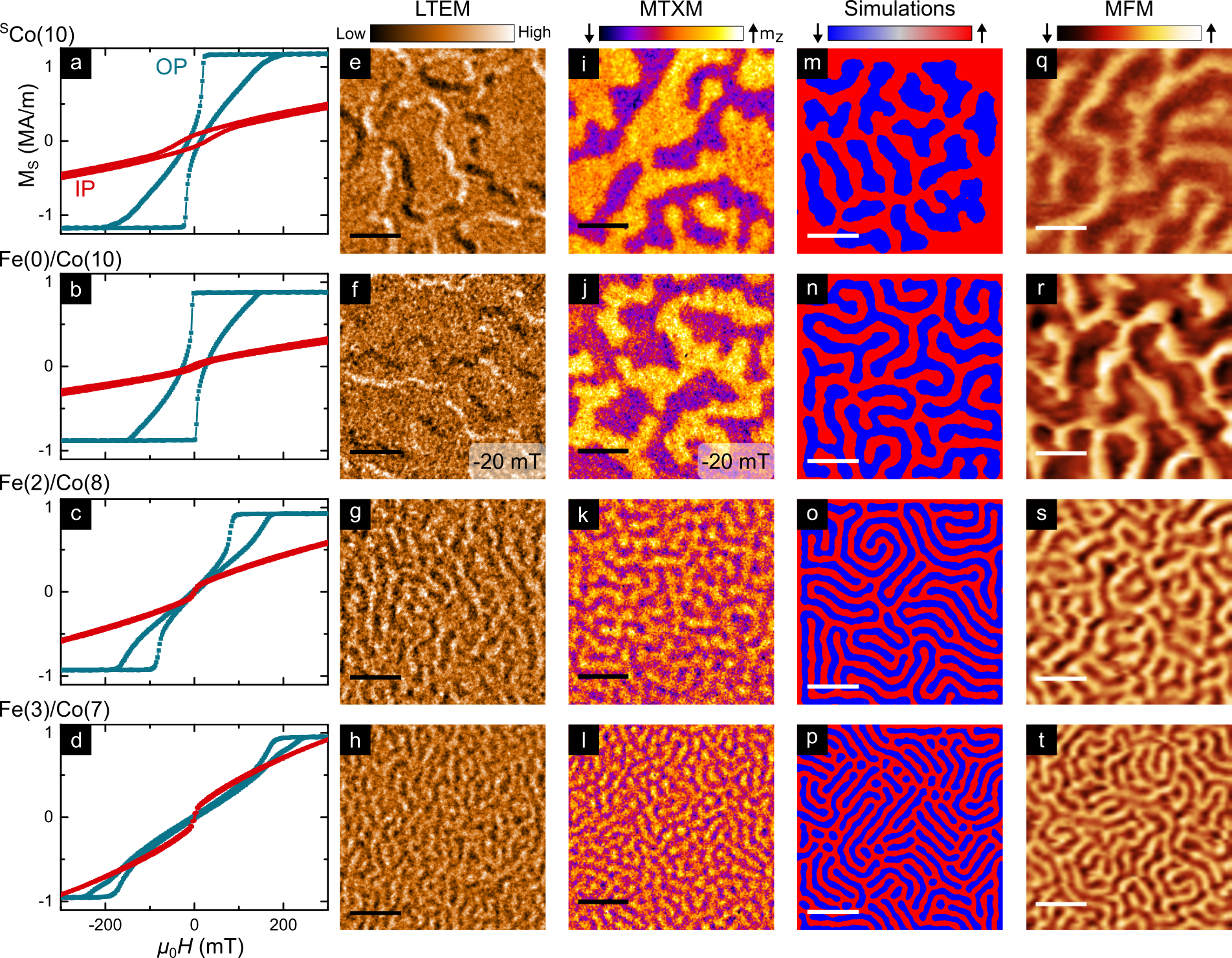}
\caption{\label{fig:overview} 
\textbf{Overview of magnetometry and magnetic imaging.} 
\textbf{(a-d)} Magnetization hysteresis loops for in-plane (IP, red) and out-of-plane (OP, teal) orientations of applied field, $\mu_0H$, across all four samples. Equilibrium magnetic configurations of all four samples as observed in \textbf{(e-h)} LTEM images  ($\alpha=20\degree$ and $-2$~mm defocus), \textbf{(i-l)} MTXM images, \textbf{(m-p)} micromagnetic simulations and \textbf{(q-t)} MFM images. All images are at the same scale and are acquired at zero-field apart than (f,j), which are at $-20$~mT. Scalebar: 0.5~$\mu$m. }
\end{figure}

The magnetic properties of the samples studied in this work are tabulated in \ref{tab:magnetic_props} and plotted in \ref{fig:magnetic_props}. The saturation magnetization, $M_\text{s}$, and effective anisotropy, $K_\text{eff}$, were determined from magnetometry measurements (see \ref{fig:overview}a-d). Following procedures established in previous works \citep{soumyanarayanan_tunable_2017}, the exchange stiffness, $A_\text{est}$, was estimated with density functional theory (DFT)\citep{bottcher_dzyaloshinskii-moriya_2020}. Meanwhile, the iDMI value, $D_\text{est}$ was estimated by comparing the periodicity of micromagnetic simulations at zero field with that from magnetic force microscopy (MFM) images (see \ref{fig:overview}q-t) \citep{soumyanarayanan_tunable_2017, woo_observation_2016, moreau-luchaire_additive_2016}. Notably, the values $A_\text{est}$ and $D_\text{est}$ for the 14x multilayers studied in this work are in good agreement with values obtained for corresponding 1x films from Brillouin light scattering (BLS) experiments \citep{bottcher_dzyaloshinskii-moriya_2020}. To check if a small iDMI affects the simulation of the $^\text{S}$Co(10) stack, we performed additional hysteresis loop simulations with iDMI of 0.1~mJ/m$^2$\citep{bottcher_dzyaloshinskii-moriya_2020} and found the results to be virtually identical to our original zero iDMI simulations.

\begin{table}[b] 
\begin{ruledtabular}
\begin{tabular}{llccccc}
\textbf{\text{Acronym}}&
\textbf{\text{Stack Composition}}&
\textbf{$\boldsymbol{M}_\text{s}$}&   \textbf{$\boldsymbol{K}_\text{eff}$}& \textbf{$\boldsymbol{D}_\text{est}$}& \textbf{$\boldsymbol{A}_\text{est}$}& \textbf{$\boldsymbol{\kappa}$}\\
&   &   (MA/m) & (MJ/m$^3$) &  (mJ/m$^2$)   &   (pJ/m)&\\
\colrule
\textbf{$^\text{S}$Co(10)}   &[Pt(10)/Co(10)/Pt(10)]$_{14}$      & 1.18  & 0.68 & 0   & 17.3  &0\\
\textbf{Fe(0)/Co(10)}        &[Ir(10)/Co(10)/Pt(10)]$_{14}$      & 0.88  & 0.47 & 1.3 & 17.8  &0.3\\
\textbf{Fe(2)/Co(8)}         &[Ir(10)/Fe(2)/Co(8)/Pt(10)]$_{14}$ & 0.93  & 0.22 & 1.8 & 12.8  &0.9\\
\textbf{Fe(3)/Co(7)}         &[Ir(10)/Fe(3)/Co(7)/Pt(10)]$_{14}$ & 0.96  & 0.08 & 2.0 & 13.6  &1.5\\
\end{tabular}
\end{ruledtabular}

\caption{\label{tab:magnetic_props}%
\textbf{ Magnetic properties of multilayer samples.}
List of multilayer samples used in this work, with layer thickness in angstroms in parentheses. The magnetic properties -- saturation magnetization ($M_\text{s}$), effective magnetic anisotropy  ($K_\text{eff}$), estimated iDMI ($D_\text{est}$), estimated exchange stiffness ($A_\text{est}$) as well as the stability parameter, $\kappa$ -- are tabulated.}
\end{table}

In \ref{fig:overview}, we show the magnetization hysteresis loops, $M(H)$, with applied field $\mu_0H$ in the in-plane (IP) and out-of-plane (OP) directions. We also show an overview of LTEM, MTXM and micromagnetic simulations for the four samples at zero field or low fields, in the labyrinthine stripe state. The magnetic morphology captured by the two experimental imaging techniques is consistent.  The simulation results are also largely in agreement with the experimental images.

\section{Image analysis\label{sec:image_analysis}}

The analysis of LTEM and MTXM images used in this work was performed with custom-written Python code. To remove low spatial frequency background, a duplicate convolved with a large Gaussian kernel was subtracted from the original image. To reduce the high spatial frequency noise while preserving as much information as possible, a small median filter was then applied. In cases where binarization was required, the threshold was automatically selected using the Otsu algorithm. To remove further noise, if any, small regions below a certain areal threshold were disregarded.

Manuscript Fig. 2  shows the averaged domain width from MTXM images. To obtain this result, background removal, denoising and binarization were performed as described above. Next, a Euclidean distance transform -- which yields the shortest distance from the domain edge to the medial axis -- was applied on the binarized image using built-in functions in the scikit-image library \citep{walt_scikit-image:_2014}. Finally, the average domain width was extracted as twice the average distance to the medial axis.

\begin{figure}[t]
\includegraphics[width=0.5\columnwidth]{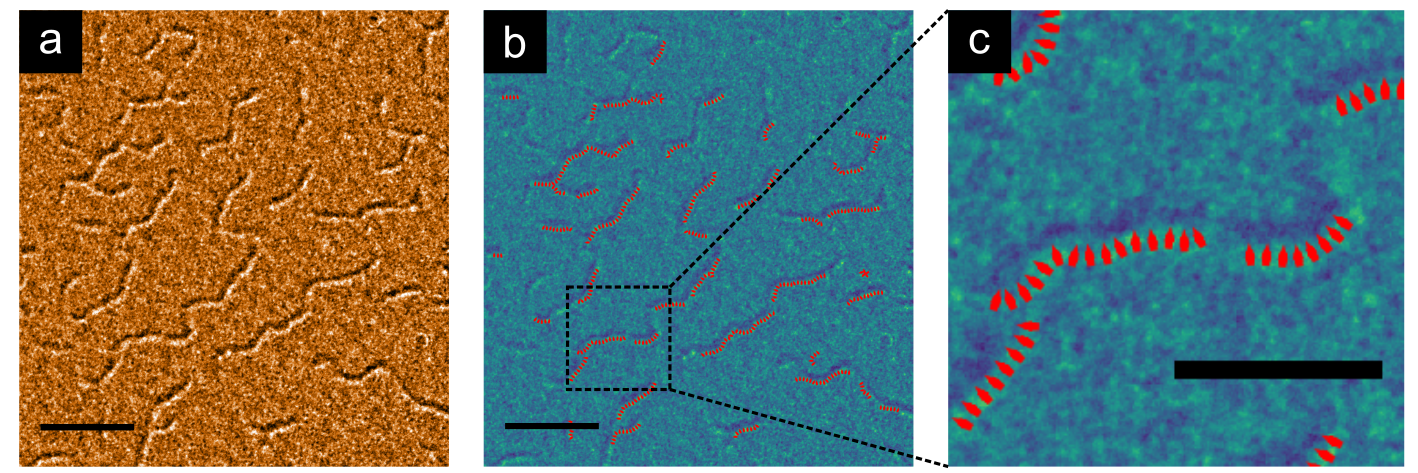}
\caption{\label{fig:linecuts_analysis} 
\textbf{Illustration of linecut analysis for LTEM images.}  
An illustration of the intermediate processing steps used to obtain linecuts from LTEM images shown in manuscript Fig. 1f-g. \textbf{(a)} Raw LTEM image of Fe(0)/Co(10) at $-100$~mT, $\alpha=-20\degree$ and $-2$~mm defocus (scalebar: 1 um). \textbf{(b)} Algorithmically determined linecuts (red lines) overlaid on the image in (a). \textbf{(c)} A zoomed-in view of (b), showing individual linecuts, as indicated by arrows (scalebar: 0.5 um).}
\end{figure}

Manuscript Fig. 1 shows linecuts across stripe domains at fixed intervals from LTEM images. To obtain these linecuts, a skeletonization operation was enacted on the LTEM images after the basic image processing and binarization described above. Next, all foreground pixels connected to more than two other pixels were removed. This ensured that there were no branches and that all regions, defined as foreground pixels that are 4-connected, are linear. Each region was then fitted to a spline, and the normal to the spline was computed. Finally, linecuts were acquired from the original, unprocessed LTEM image using methods in scikit-image library, at regular intervals on the spline. Linecuts exceeding the boundaries of the image were excluded. The position and the direction of the algorithmically chosen linecuts are shown in \ref{fig:linecuts_analysis}. The acquired linecuts were further binned by angle, and only those within $10\degree$ from the tilt axis were included in calculating the averaged output. \\

For analysis of LTEM images acquired at varying tilt angles, an additional preprocessing step was required. The sequence of images had to be aligned to the first image, as the latter was used to determine the position of each linecut. The image alignment was done manually with ImageJ by selecting about 15 corresponding landmarks on each image and calculating an affine transformation to map each image in the tilt sequence to the first image. Manuscript Fig. 1 does not show the linecut for zero tilt of Fe(0)/Co(10). This is because, in the absence of visible magnetic textures at zero tilt, the alignment landmarks for that image could not be located. Similar image sequence alignment issues precluded the use of this method for analyzing Fe(3)/Co(7) images (see \ref{sec:ltem}).

\begin{figure}[t]
\includegraphics[width=0.4\columnwidth]{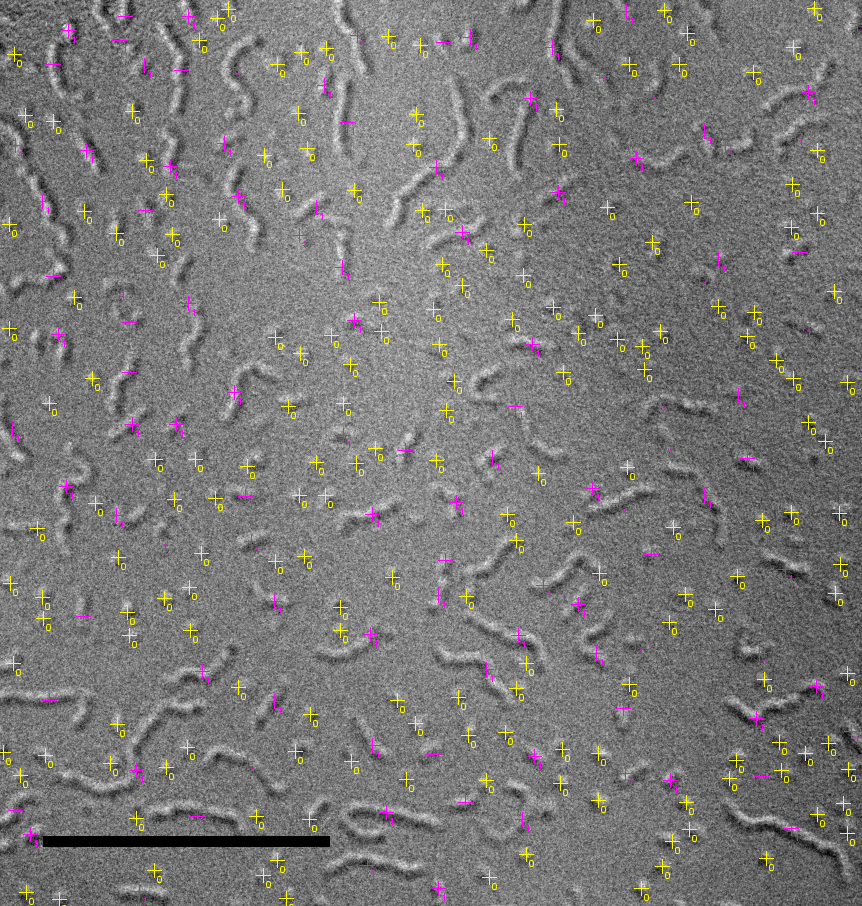}
\caption{\label{fig:counting} 	  
\textbf{Identification of textures in LTEM images.} 
A representative LTEM image of Fe(2)/Co(8) at 150~mT, $\alpha=20\degree$ and $-2$~mm defocus (scalebar: 1 um) used to demonstrate the counting of observed skyrmions and stripes. Skyrmions are indicated with yellow crosses, and stripes with pink crosses.}
\end{figure}

Manuscript Fig. 4 details the statistics of skyrmion and stripe number densities with varying magnetic field, which were obtained as follows. For micromagnetic simulations (manuscript Fig. 4c,d), algorithmic counting gave reliable results after denoising and binarization due to the high signal-to-noise ratio (SNR) of domains in simulated images. Domains that exceeded a visually determined circularity threshold of 0.65, were classified as skyrmions, while the remainder otherwise, were counted as stripes. Here, circularity is defined as $4\pi A/P^2$, where $A$ and $P$ are the area and perimeter of the domain respectively. For LTEM, attempts to use such algorithmic counting of stripes and skyrmions did not give results consistent with visual inspection. Hence, we visually identified every texture in LTEM images as a skyrmion or stripe. A typical LTEM image of Fe(2)/Co(8) with skyrmions and stripes identified as above is shown in \ref{fig:counting}. 

\section{LTEM Imaging\label{sec:ltem}}

\begin{figure}[t]
\includegraphics[width=0.6667\columnwidth]{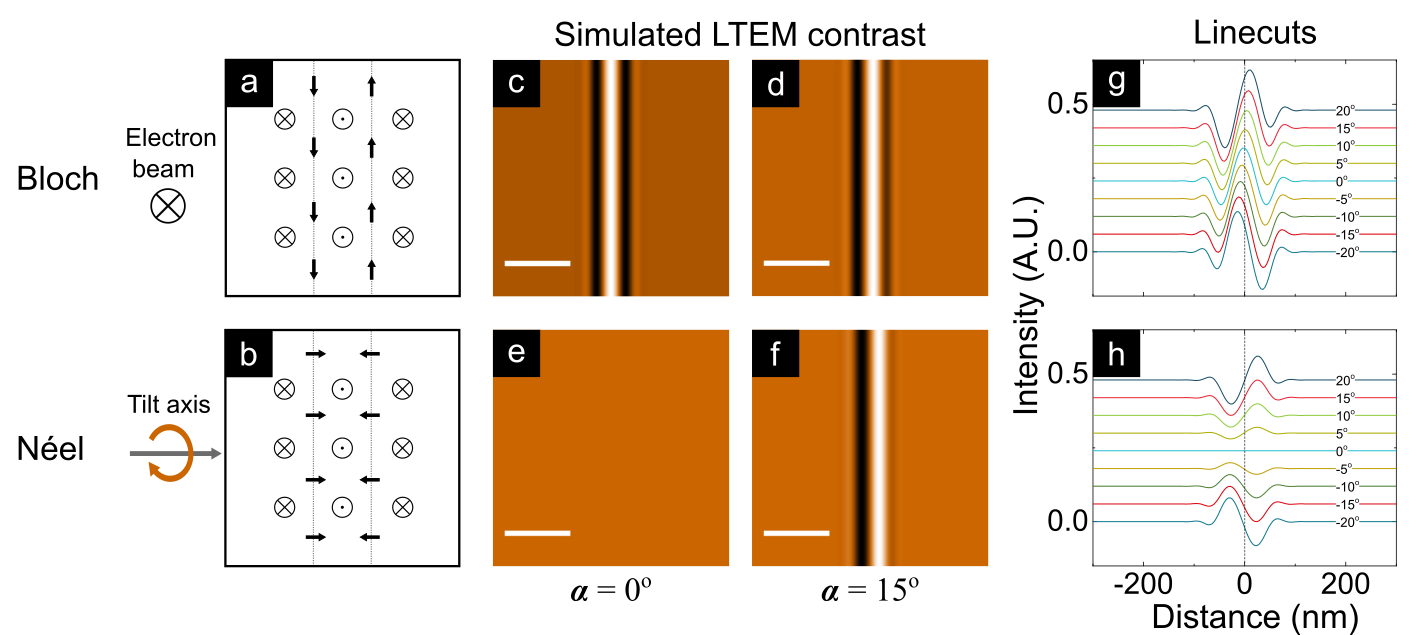}
\caption{\label{fig:malts} 
\textbf{Simulated LTEM contrast.} 
\textbf{(a-b)} Schematic of an artificially constructed magnetic stripe for MALTS simulations with Bloch and Néel DWs respectively. Vector symbols represent the direction of the electron beam and local magnetization. \textbf{(c-f)} MALTS simulated LTEM contrast at $\alpha=0\degree$ and $15\degree$ (scalebar: 200~nm). \textbf{(g-h)} Linecuts across the simulated LTEM images with $\alpha$ varying over $\pm$ 20$\degree$.}
\end{figure}

In this work, spatial evolution of LTEM contrast of textures has been used to quantify their helicity. As LTEM contrast is not  intuitive to non-experts, we show in \ref{fig:malts} the simulated LTEM contrast -- generated using an open-source software, MALTS\citep{walton_malts:_2012} -- for an idealized magnetic stripe with Bloch and Néel DWs. Also shown are the linecuts perpendicular to the domain for various tilt angles (c.f. manuscript Fig. 1f-g). At zero tilt, the Bloch DW pair produces contrast symmetric about the center of the stripe, whereas the Néel DW pair give zero contrast. In both cases, an additional antisymmetric contrast is present for finite tilt.

\begin{figure}[t]
\includegraphics[width=0.45\columnwidth]{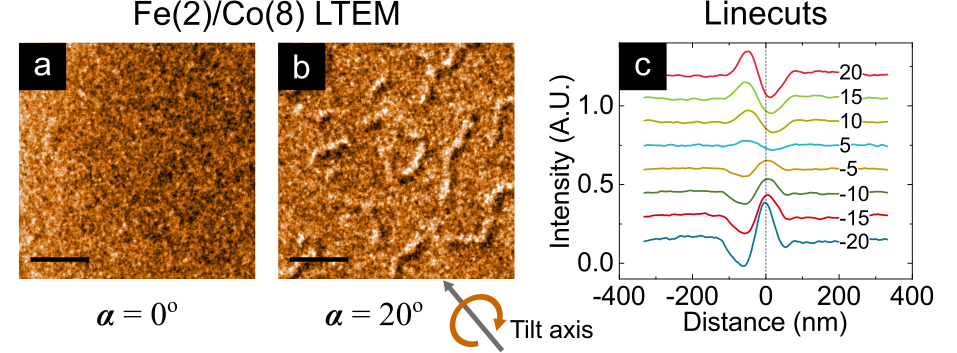}
\caption{\label{fig:ltem_additional} 
\textbf{LTEM analysis of DW helicity for Fe(2)/Co(8).}  
\textbf{(a-b)} LTEM images of Fe(2)/Co(8) sample at $\alpha=0\degree$ and $20\degree$, at a field of 100~mT and at $-2$~mm defocus (scalebar: 0.5~$\mu$m). \textbf{(c)} Averaged cross-sectional linecuts determined over the experimental field-of-view with tilt angle varying over $\pm 20 \degree$ (c.f. manuscript Fig. 1d-e).}
\end{figure}

In manuscript Fig. 1, we have shown tilt-dependent LTEM data for $^\text{S}$Co(10), confirming Bloch helicity of its DWs, and for Fe(0)/Co(10) -- wherein DWs have Néel helicity. In \ref{fig:ltem_additional}, we show for comparison LTEM data for  Fe(2)/Co(8), which also has, unsurprisingly, Néel DWs. Finally, the linecut extraction technique used for the previous three samples could not be applied to Fe(3)/Co(7).  In this case, owing to the higher texture density,  images with varying tilt could not be reliably aligned (see \ref{sec:image_analysis}). Nonetheless, a visual inspection of the contrast evolution with tilt suggests that the textural helicity for Fe(3)/Co(7) is very similar to that of Fe(2)/Co(8) hosting Néel DWs. 

\begin{figure}[t]
\includegraphics[width=0.3333\columnwidth]{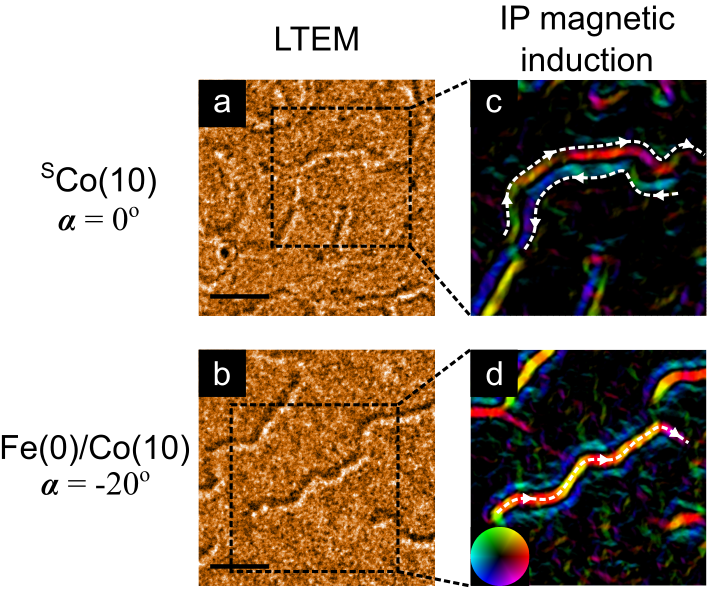}
\caption{\label{fig:tie} 
\textbf{Transport-of-intensity equation (TIE) analysis of LTEM images.}  
\textbf{(a,b)} LTEM images of  $^\text{S}$Co(10) and Fe(0)/Co(10) at $\alpha=0\degree$ and $-20\degree$ respectively  (scalebar: 0.5~$\mu$m). \textbf{(c,d)} Reconstructed IP magnetic induction of $^\text{S}$Co(10) and Fe(0)/Co(10) using LTEM images acquired at defoci of $\pm2$~mm. The magnetization direction is represented by color. Inset to (d) shows color wheel, and white dotted lines with arrows are guides for the eye.}
\end{figure}

The transport-of-intensity equation (TIE) has previously been used to reconstruct the magnetization of Bloch and Néel textures with varying degrees of success \citep{pollard_observation_2017}. Correspondingly in \ref{fig:tie}, we show the IP magnetic induction for $^\text{S}$Co(10) and Fe(0)/Co(10), each reconstructed using two LTEM images at defoci of $\pm2$~mm using a TIE-based commercial software, QPt$^\text{TM}$ for Digital Micrograph from HREM Research, Japan (c.f. manuscript Fig. 1). For $^\text{S}$Co(10), we can clearly see the evidence for Bloch DWs  in \ref{fig:tie}c. In contrast, for Fe(0)/Co(10), due to the finite tilt angle, a part of the out-of-plane magnetization of the textures are projected in-plane and is visible in \ref{fig:tie}d. Crucially, no evidence of Bloch DWs is detectable for Fe(0)/Co(10).

\begin{figure}[t]
\includegraphics[width=0.55556\columnwidth]{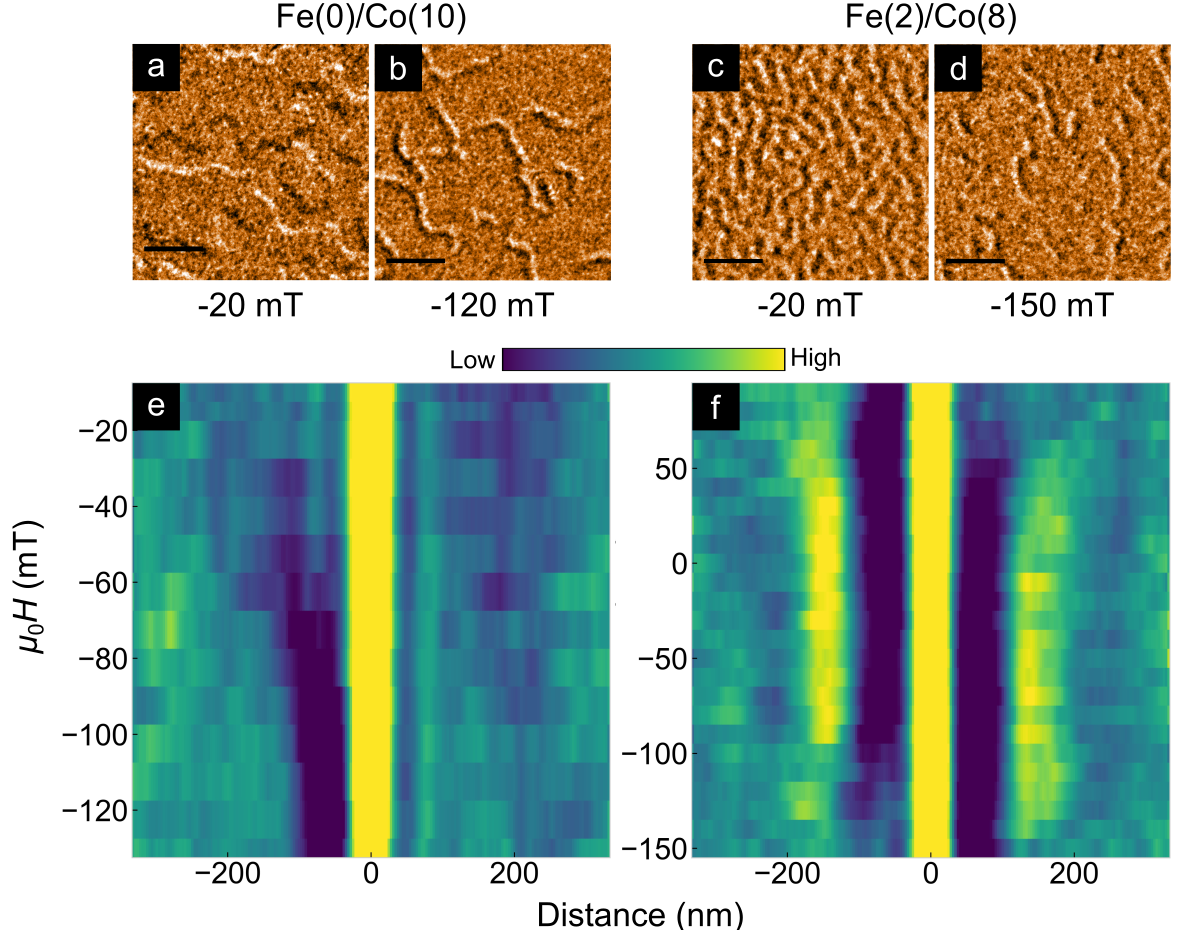}
\caption{\label{fig:ltem_compressibility}  
\textbf{Domain compressibility investigated by LTEM.} 
\textbf{(a-d)} Representative LTEM images of Fe(0)/Co(10) and Fe(2)/Co(8) at low and high magnetic fields (scalebar: 0.5~$\mu$m). \textbf{(e,f)} Stacked color plots of field-of-view averaged cross-sectional linecuts for Fe(0)/Co(10) and Fe(2)/Co(8) as a function of varying magnetic field (along y-axis). color represents LTEM intensity along the linecut. }
\end{figure}

In manuscript Fig. 2, we present MTXM results as direct experimental evidence when investigating domain compressibility as a function of $\kappa$. The same trend could be observed indirectly with LTEM. To this end, LTEM imaging was performed along a hysteresis loop, as shown in \ref{fig:ltem_compressibility}a-d. It is visually apparent that at low fields of $-20$~mT, the average domain width is much greater for Fe(0)/Co(10) (\ref{fig:ltem_compressibility}a) than that of Fe(2)/Co(8) (\ref{fig:ltem_compressibility}c). However, at higher fields, their domain widths are comparable (\ref{fig:ltem_compressibility}b,d). This visual observation could be placed on quantitative standing if we employ the previously established algorithmic linecut analysis, but this time as a function of the magnetic field, as shown in \ref{fig:ltem_compressibility}e,f (c.f. MTXM in manuscript Fig. 2i). As the linecuts are centered on the intensity peaks, the domain width is simply the separation between the intensity peak (bright) and trough (dark) regions. For Fe(0)/Co(10) at low fields, the trough is separated $\sim$200~nm from the peak and is heavily smeared. This is because DWs are well separated (i.e. domains are wide) with weak spatial correlation. With increasing field, the trough moves towards the peak (i.e. domain width shrinks) and becomes more correlated. This is consistent with the highly compressible behavior seen in MTXM data. In contrast, for Fe(2)/Co(8), the peaks and troughs are always adjacent and have strong spatial correlation at all fields. This directly supports our claim that the domains are incompressible for Fe(2)/Co(8).

\section{Micromagnetic Simulations}
Micromagnetic simulations were performed with MuMax3, using the magnetic parameters listed in \ref{tab:magnetic_props}\citep{vansteenkiste_design_2014}.

In the manuscript, we discuss the lack of evidence for hybrid chirality in Fe(0)/Co(10), Fe(2)/Co(8) and Fe(3)/Co(7) based on our LTEM data. Our micromagnetic simulations -- which will be elaborated here -- are consistent with our experiments in suggesting that hybrid chirality, if present at all, is very limited in these three samples. \\

\begin{figure}[t]
\includegraphics[width=0.75\columnwidth]{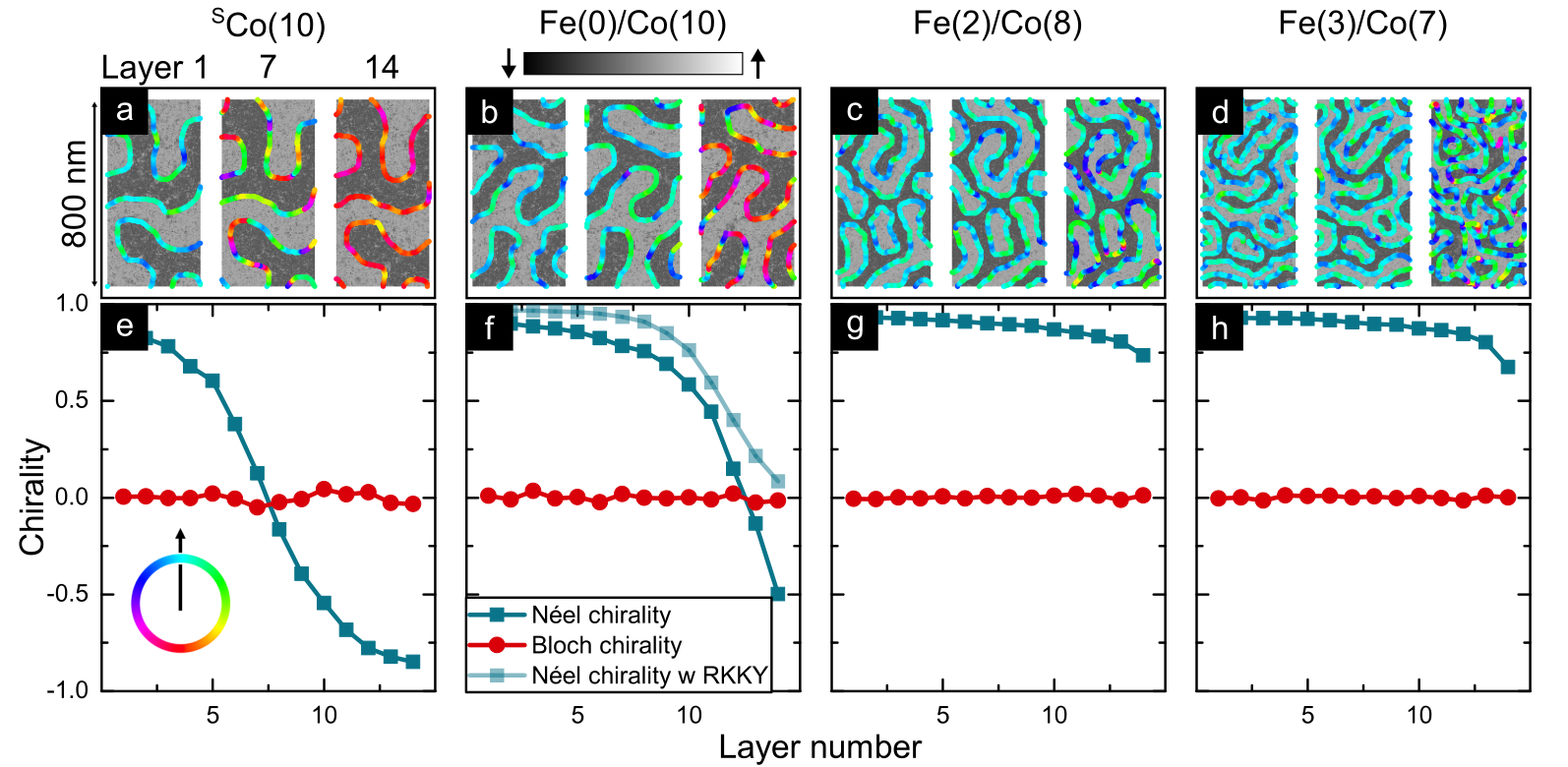}
\caption{\label{fig:sim_helicity} 	  
\textbf{Layer-dependent chirality of DWs from simulations.} 
\textbf{(a-d)} Representative cropped regions for layers 1, 7 and 14 from zero-field micromagnetic simulations of the four samples studied in this work. The grayscale colormap represents the OP magnetization, while colored ribbons show the IP magnetization orientation of the DWs relative to the DW normal vector. Inset in (e) shows the color wheel for IP orientation of DW magnetization, wherein the black arrow represents the direction of the DW normal. \textbf{(e-h)} The spatially averaged degree of Néel and Bloch chiralities as a function of layer number for the four samples. Here, the degree of Néel and Bloch chiralities are defined as $\mathbf{n}_\text{DW} \cdot \mathbf{m}$ and $\mathbf{n}_\text{DW} \times \mathbf{m}$ ($\mathbf{n}_\text{DW}$ is the DW normal unit vector and $\mathbf{m}$ is the magnetization unit vector) respectively. The light teal line plot in (f) shows Bloch chirality when interlayer RKKY coupling -- at 20\% of direct exchange -- is considered.}
\end{figure}

The zero-field micromagnetic simulations were analyzed and DWs regions for each layer were isolated for analysis. \ref{fig:sim_helicity}a-d shows the OP magnetization (grayscale) and IP orientation of the DW magnetization relative to the DW normal vector (colored ribbons). The dot and cross products of the magnetization unit vector with the normal unit vector of the DWs were computed, which can be identified as the degree of Néel and Bloch chiralities respectively. The averaged layer-dependent Néel and Bloch chiralities are plotted as a function of layer number for the four samples in \ref{fig:sim_helicity}e-h. \\

Our simulations suggest that $^\text{S}$Co(10) ($D_\text{est}=0$) has considerable hybrid chirality \citep{legrand_hybrid_2018}. The layers near the center have Bloch helicity, while the layers near the top and bottom form Néel caps, consistent with previous works \citep{legrand_modeling_2018, legrand_hybrid_2018}. However, upon considering all layers as a whole, both Bloch and Néel chiralities average to zero. Therefore, we label this sample as achiral \citep{montoya_tailoring_2017, yu_magnetic_2012}. In comparison, the other three samples -- Fe(0)/Co(10), Fe(2)/Co(8) and Fe(3)/Co(7) -- with $D_\text{est}>0$ are strongly chiral, i.e. have fixed handedness. Even if a Bloch center is present in Fe(0)/Co(10) ($D_\text{est}= 1.3$~mJ/m$^2$), it might only exist in one or two layers which is difficult to observe experimentally. Moreover, we show that Fe(0)/Co(10) is at the threshold of hybrid chirality since the Bloch center vanishes if we consider a small but finite interlayer RKKY coupling -- amounting to 20\% of $A_\text{est}$ -- as shown in \ref{fig:sim_helicity}f \citep{legrand_room-temperature_2019}. Hence, within this work, our simulations demarcate the boundaries of hybrid chirality as $\kappa \lesssim 0.3$. Meanwhile, we do not expect a Bloch center in Fe(2)/Co(8) and Fe(3)/Co(7) ($D_\text{est} \geq 1.8$~mJ/m$^2$), since their Néel chirality did not change sign across layers.

\begin{figure}[t]
\includegraphics[width=0.5\columnwidth]{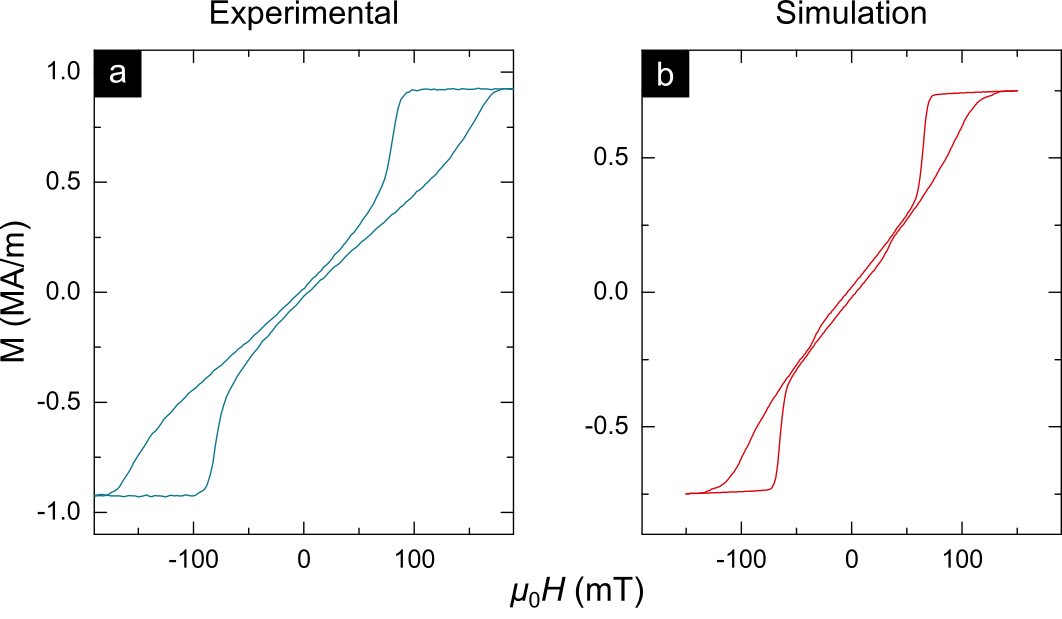}
\caption{\label{fig:sim_MH} 
\textbf{Hysteresis loop simulation.}  
Comparison of the (a) experimental and (b) simulated OP hysteresis loop for sample Fe(2)/Co(8). Hysteresis loop simulations were performed with MuMax3 using procedures described in the text.}
\end{figure}

Hysteresis loops are simulated by time-stepping in the presence of a sweeping applied magnetic field\citep{vansteenkiste_design_2014}. The rate of field sweep is approximately 10$^6$~T/s due to computational constraints, and so an entire hysteresis loop simulation is swept in about 600~ns. In order to ensure that simulated magnetization configurations may cross energy barriers at rates corresponding to conventional magnetometry experiments despite these constraints, the simulation temperature needs to be correspondingly elevated (e.g. to 850 -- 900~K). A typical hysteresis loop simulation result is shown in \ref{fig:sim_MH} for direct comparison with experiments. Notwithstanding quantitative discrepancies of $\sim$20\% in saturation field and magnetization, the simulated hysteresis loops fully reproduce the key experimental features, such as the sheared shape and various kinks.

\begin{figure}[t]
\includegraphics[width=0.5\columnwidth]{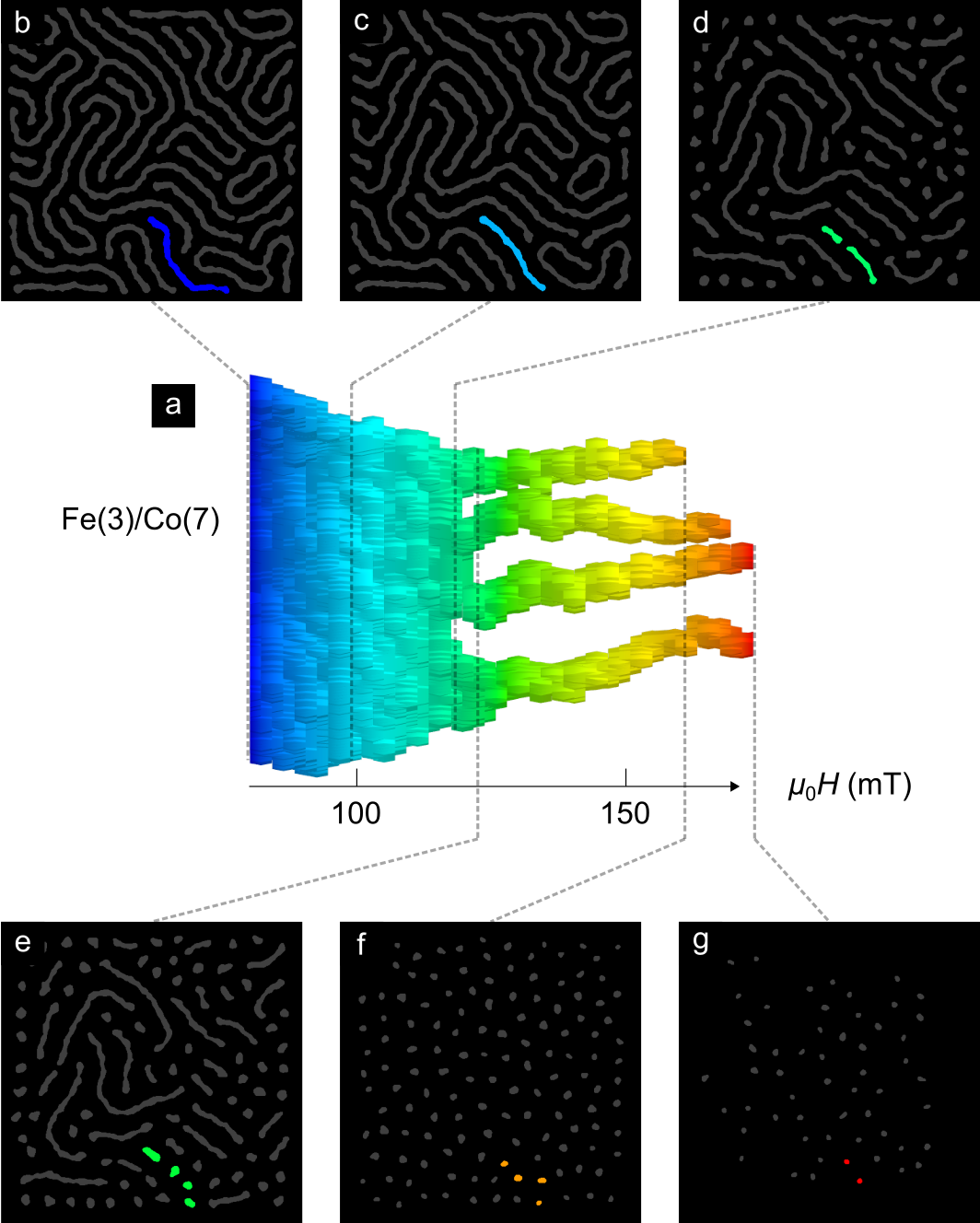}
\caption{\label{fig:evolution_map} 
\textbf{Construction of a field evolution map of a typical domain.}  
\textbf{(a)} Field evolution map of a typical Fe(3)/Co(7) domain as shown in manuscript Figure 4b. Color denotes the magnitude of the magnetic field. \textbf{(b-g)} Simulated magnetization images (field-of-view: 2~$\mu$m) at various fields that are combined to construct (a). The chosen domain(s) is (are) highlighted in each slice.}
\end{figure}

It is convenient to visualize skyrmion formation from magnetic stripes in simulations with a “field evolution map” of a typical magnetic domain. This is created by stacking the 2D footprint of a typical domain across varying fields into a 3D object, where the third axis is the magnetic field. This process is illustrated in \ref{fig:evolution_map}, where the chosen domain is in color. This 3D object is then rotated such that the field axis is horizontal, as shown in manuscript Fig. 4a,b. The mapping of the chosen domain across image slices in the magnetic field sweep is established using their spatial overlap -- since the domains do not move appreciably within one field step. 

We also include two supplementary videos showing the evolution of the magnetic textures as a function of applied field that is used in the construction of manuscript Fig. 4(a,b).

\textbf{Video 1:} This video shows the simulated evolution of magnetic textures of Fe(2)/Co(8) over a 2~$\mu$m field-of-view from 72~mT to 132~mT. Highlighted domains denote the representative evolution of stripes to skyrmions for Fe(2)/Co(8), and are horizontally stacked across fields to form manuscript Fig. 4(a).

\textbf{Video 2:} This video shows the simulated evolution of magnetic textures of Fe(3)/Co(7) over a 2~$\mu$m field-of-view from 80~mT to 170~mT. Highlighted domains denote the representative evolution of stripes to skyrmions for Fe(3)/Co(7), and are horizontally stacked across fields to form manuscript Fig. 4(b).

\section{Geodesic Nudged Elastic Band (GNEB) simulations}

GNEB atomistic calculations were performed with the Fidimag package \citep{bisotti_fidimag_2018}. A single layer of magnetic spins arranged in a two-dimensional square lattice with a cell size of 1~nm $\times$ 1~nm was used for all the GNEB simulations.

GNEB is a well-established method to calculate the energy barrier for a transition between two fixed metastable magnetic configurations, under the constraint of fixed magnetic moments magnitude \citep{bessarab_method_2015}. Since GNEB is an atomistic calculation, the energies of the configurations are calculated in accordance with the atomistic Heisenberg Hamiltonian, taking into account exchange, DMI, uniaxial anisotropy and magnetostatic interactions \citep{cortes-ortuno_thermal_2017}. To estimate the energy barriers of the stripe-skyrmion fission process, the initial metastable configuration was chosen to be single stripe domain relaxed at $-0.04$ mT, and the final configuration to be two metastable skyrmions relaxed at the same external field. A total of 16 intermediate transition states and their energies were calculated by the GNEB algorithm, from which the energy barrier of the process can be estimated by interpolation. The resulting energy barriers of the stripe-skyrmion transition for Fe(2)/Co(8) and Fe(3)/Co(7) are shown in the main text.

\section{Compressibility}

\begin{figure}[t]
\includegraphics[width=0.5\columnwidth]{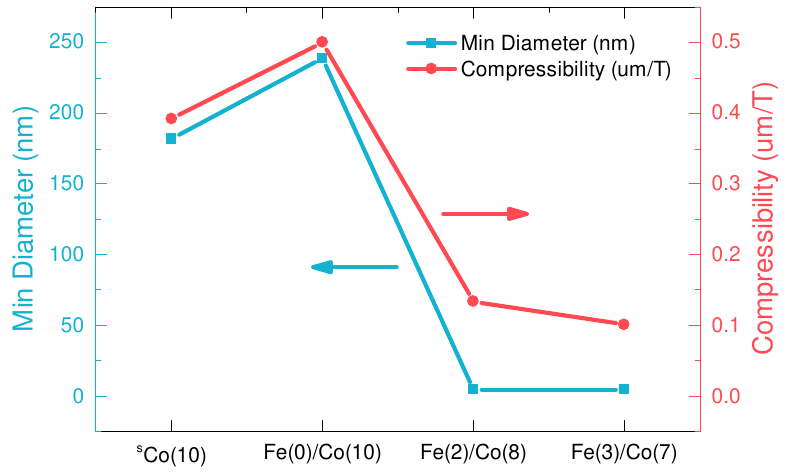}
\caption{\label{fig:min_sk_diameter}  
\textbf{Minimum skyrmion diameter predicted from isolated skyrmion model.} 
Theoretical results obtained by feeding in magnetic parameters listed  \ref{tab:magnetic_props} into model of isolated skyrmions \citep{buttner_theory_2018}. This is compared with experimentally measured compressibility from manuscript Fig. 2k. The trends of both metrics are in close agreement -- a sharp divide separates bubble skyrmions ($^\text{S}$Co(10), Fe(0)/Co(10)) from compact skyrmions (Fe(2)/Co(8), Fe(3)/Co(7)).}
\end{figure}

Here, we compare two metrics that can be used to differentiate bubble skyrmions and compact skyrmions. In \ref{fig:min_sk_diameter}, we show the compressibility of stripes reproduced from manuscript Fig. 2k and minimum diameter of isolated skyrmions \citep{buttner_theory_2018} for the four samples in this work. Both metrics show a clear distinction between $^\text{S}$Co(10), Fe(0)/Co(10) (bubble skrymions) and Fe(2)/Co(8), Fe(3)/Co(7) (compact skyrmions).

\begin{figure}[t]
\includegraphics[width=0.5\columnwidth]{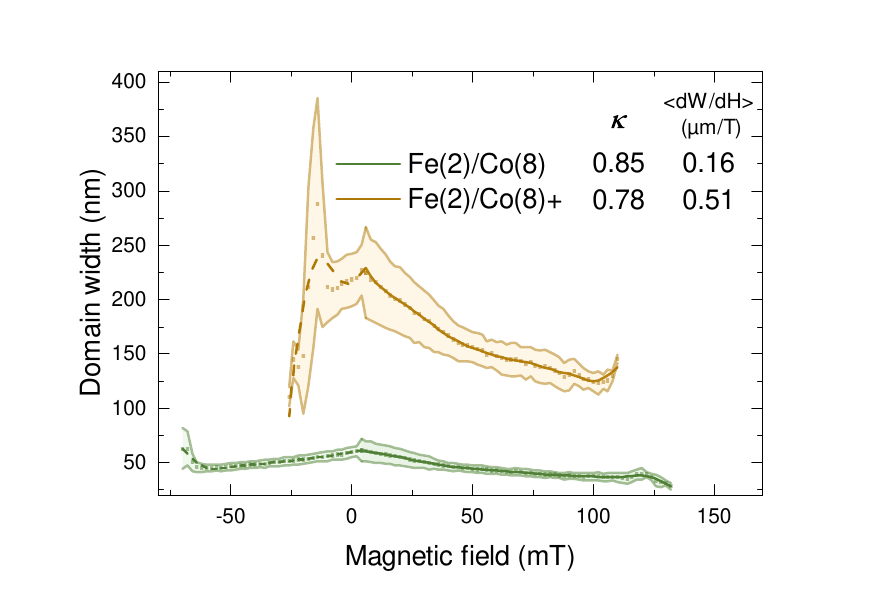}
\caption{\label{fig:26jSimOriginalVsAexBoost}  
\textbf{Simulated domain width field evolution of Fe(2)/Co(8) and Fe(2)/Co(8)+.} 
Micromagnetic simulations of domain width as functions of applied magnetic field of Fe(2)/Co(8) and Fe(2)/Co(8)+, a fictitious sample with identical magnetic parameters as Fe(2)/Co(8) except an exchange stiffness that is  20\% larger than Fe(2)/Co(8). Inset lists the $\kappa$ and $\langle dW/dH \rangle$ values for both samples.}
\end{figure}

One might notice a correlation between domain compressibility, $\langle dW/dH \rangle$ and $K_\text{eff}$. We believe that this correlation arises from the fact that $\kappa=\pi D/4\sqrt{AK_\text{eff}}$, and therefore $K_\text{eff}$ is a component of $\kappa$. Here, we show with micromagnetic simulations that $K_\text{eff}$ cannot fully account for the domain compressibility trends. We study a fictitious sample, labeled as Fe(2)/Co(8)+, which has identical magnetic parameters as Fe(2)/Co(8) –- including $K_\text{eff}$ -- except a direct exchange, $A_\text{est}$, that is 20\% larger, which results in a smaller $\kappa$. We simulate the hysteresis loop of Fe(2)/Co(8)+, extract the domain width and compute the $\langle dW/dH \rangle$ using the same protocol as detailed in the manuscript. The comparison between Fe(2)/Co(8) and Fe(2)/Co(8)+ is shown in \ref{fig:26jSimOriginalVsAexBoost}. As can be seen, $\langle dW/dH \rangle$ can vary considerably even if $K_\text{eff}$ is kept constant across samples.

\noindent \begin{center}
{\small{}\rule[0.5ex]{0.6\columnwidth}{0.5pt}}{\small\par}
\par\end{center}

\end{document}